\documentclass[amsmath,amssymb,aps,prb,twocolumn,showpacs]{revtex4}

\usepackage{graphicx}
\usepackage{amsmath,latexsym}
\usepackage{dcolumn}
\usepackage{bm}
\usepackage{sidecap}
\usepackage{multirow}
\usepackage{bm}

\usepackage[usenames,dvipsnames]{color}
\newcommand{\green}{\textcolor{OliveGreen}}
\newcommand{\red}{\textcolor{Red}}

\def\jhm#1{\red{[JH: #1]}}
\def\dvm#1{\green{[DV: #1]}}

\newcommand{\mylabel}[1]{\blue{\small\hbox{\;#1}}\label{#1}} 

\def\dvm#1{}
\def\jhm#1{}
\renewcommand{\mylabel}[1]{\label{#1}}

\def\n{\noindent}

\def\bea{\begin{eqnarray}}
\def\nn{\nonumber\\}
\def\eea{\end{eqnarray}}
\def\beq{\begin{equation}}
\def\eeq{\end{equation}}

\def\BTO{BaTiO$_3$}
\def\BZO{BaZrO$_3$}
\def\STO{SrTiO$_3$}
\def\PTO{PbTiO$_3$}
\def\BSTO{(Ba$_{0.5}$Sr$_{0.5}$)TiO$_3$}

\newcommand{\equ}[1]{Eq.~(\ref{eq:#1})}
\newcommand{\equs}[2]{Eqs.~(\ref{eq:#1}) and (\ref{eq:#2})}
\newcommand{\equr}[2]{Eqs.~(\ref{eq:#1}-\ref{eq:#2})}

\def\P{{\bf P}}
\def\E{{\cal E}}

\def\rr{{\bf r}}
\def\u{{\bf u}}
\def\R{{\bf R}}

\def\k{{\bf k}}

\def\be{{\boldsymbol \eta}}
\def\bn{{\boldsymbol \nu}}

\def\a{\alpha}
\def\b{\beta}
\def\c{\gamma}
\def\d{\delta}
\def\e{\eta}
\def\n{\nu}
\def\nb{\nu^{\rm s}}

\def\m{\mu}
\def\t{\tau}
\def\l{\lambda}
\def\str{\epsilon}

\def\mlone{\mu_\textrm{L1}}
\def\mltwo{\mu_\textrm{L2}}
\def\mtr{\mu_\textrm{T}}

\def\blone{_\textrm{L1}}
\def\bltwo{_\textrm{L2}}
\def\blt{_\textrm{T}}

\def\pD{^D}
\def\pE{^\E}

\def\me{\m\pE}
\def\md{\m\pD}

\def\mbare{\m^{\textrm{el},\E}}
\def\mbard{\m^{\textrm{el},D}}

\def\mpri{\m'}

\def\Te{T\pE}
\def\Td{T\pD}
\def\Ke{K\pE}
\def\Kd{K\pD}
\def\Ze{Z\pE}
\def\Zd{Z\pD}

\def\Ki{(K^{-1})}

\def\cost{{\rm cos} \, \theta}
\def\sint{{\rm sin} \, \theta}

\def\G{\Gamma}
\def\L{\Lambda}

\def\pa{\partial}

\def\pzer{^{(0)}}
\def\pone{^{(1)}}
\def\ptwo{^{(2)}}
\def\pthr{^{(3)}}

\def\poneL{^{(1,\textrm{L})}}
\def\poneT{^{(1,\textrm{T})}}
\def\ptwoL{^{(2,\textrm{L})}}
\def\ptwoT{^{(2,\textrm{T})}}

\def\bib{_{I,\,\b}}

\def\bit{_{I\tau}}
\def\biat{_{I\a\tau}}
\def\biatb{_{I\a\tau\b}}
\def\biatbc{_{I\a\tau\b\c}}
\def\bjtp{_{J\tau'}}

\def\bjtpbc{_{J\tau'\b\c}}
\def\bjtpbcd{_{J\tau'\b\c\d}}
\def\ba{_{\a}}
\def\bb{_{\b}}
\def\bc{_{\c}}
\def\bd{_{\d}}

\def\biab{_{i\a,\b}}

\def\bl{_{1111}}
\def\btm{_{1221}}
\def\btr{_{1122}}

\def\bbc{_{\b\c}}
\def\bbcd{_{\b\c\d}}
\def\biab{_{I,\,\a\b}}
\def\biabc{_{I,\,\a\b\c}}
\def\biabd{_{I,\,\a\b\d}}
\def\biabcd{_{I,\,\a\b\c\d}}

\def\bitbc{_{I\t\b\c}}
\def\bitbd{_{I\t\b\d}}
\def\bitbcd{_{I\t\b\c\d}}

\def\bjit{_{j,I\t}}

\def\bitj{_{I\t,j}}
\def\Ai{(A^{-1})}

\def\Qt{\widetilde{Q}}
\def\Lt{\widetilde{\L}}
\def\Tt{\widetilde{T}}
\def\Kt{\widetilde{K}}
\def\Gt{\widetilde{\G}}
\def\Nt{\widetilde{N}}

\def\bab{_{\a\b}}
\def\bba{_{\b\a}}
\def\babc{_{\a\b\c}}
\def\bbac{_{\b\a\c}}
\def\babcd{_{\a\b\c\d}}
\def\bbcd{_{\b\c\d}}

\def\Sr{\mathrm{Sr}}
\def\Ti{\mathrm{Ti}}
\def\O{\mathrm{O}}

\def\half{\frac{\textstyle{1}}{\textstyle{2}}}

\def\GOF{\Gamma_{15}}
\def\GTF{\Gamma_{25}}

\def\Tt{\widetilde{T}}
\def\ft{\tilde{f}}

\def\kti{\tilde{k}^{-1}}

\def\Vc{V_{\rm c}}

\def\pld{^{\rm l\hspace{0.4pt}d}}      

\def\plq{^{\rm l\hspace{0.4pt}q}}      
\def\pel{^{\rm e\hspace{0.2pt}l}}      
\def\plqT{^{\rm l\hspace{0.4pt}q,T}}      
\def\pelT{^{\rm e\hspace{0.2pt}l,T}}      
\def\plqJ{^{\rm l\hspace{0.4pt}q,J}}      
\def\pelJ{^{\rm e\hspace{0.2pt}l,J}}      
\def\pldE{^{\rm l\hspace{0.4pt}d,\E}}      
\def\pelE{^{\rm e\hspace{0.2pt}l,\E}}      
\def\pelD{^{\rm e\hspace{0.2pt}l,D}}      
\def\ppel{^{\prime\,\rm e\hspace{0.2pt}l}}      

\def\gtil{\tilde{g}}

\def\pext{^{\rm ext}}

\def\Jw{J^{[w]}}
\def\ket#1{\vert#1\rangle}
\def\bra#1{\langle#1\vert}
\def\ip#1#2{\langle#1\vert#2\rangle}

\def\PP{\mathcal{P}}

\begin{document}

\title{First-principles theory and calculation of flexoelectricity}

\author{Jiawang Hong}
\email{hongjw10@physics.rutgers.edu}
\affiliation{ Department of Physics and Astronomy, Rutgers University,
 Piscataway, NJ 08854-8019, USA }

\author{David Vanderbilt}
\affiliation{ Department of Physics and Astronomy, Rutgers University,
 Piscataway, NJ 08854-8019, USA }

\date{\today}

\begin{abstract}

We develop a general and unified first-principles theory of
piezoelectric and flexoelectric tensor, formulated in such a way
that the tensor elements can be computed directly in the context
of density-functional calculations, including electronic and
lattice contributions. We introduce a practical supercell-based
methods for calculating the flexoelectric coefficients from first
principles, and demonstrate them by computing the coefficients
for a variety of cubic insulating materials including C, Si, MgO,
NaCl, CsCl, BaZrO$_3$, BaTiO$_3$, PbTiO$_3$ and SrTiO$_3$.

\end{abstract}


\maketitle

\section{Introduction}
\mylabel{sec:intro}

Flexoelectricity (FxE) describes the linear coupling between
electric polarization and a strain gradient, and is always
symmetry-allowed because a strain gradient automatically breaks the
inversion symmetry. This is unlike the case of piezoelectricity
(coupling of polarization to strain), which arises only in
noncentrosymmetric materials.  FxE was theoretically
proposed about 50 years ago,\cite{kogan-spss64} and was
discovered experimentally four years later by
Scott\cite{scott-jcp68} and Bursian \textit{et al.}\cite{bursian-spss67}
The FxE effect received very little attention for decades because
of its relatively weak effects. Recently it has attracted increasing
attention, however, largely stimulated by the work of Ma and
Cross\cite{ma-apl01,ma-apl01-1,ma-apl02,ma-apl03,ma-apl05,ma-apl06,cross-jms06}
in which they found that the flexoelectric coefficient (FEC) could
have an order of magnitude of $\mu$C/m, three orders larger than previous
theoretical estimations.\cite{kogan-spss64}

A second reason for the revival of interest in FxE is that strain
gradients are typically much larger at the nanoscale than at
macroscopic scales.  For example, a 1\% strain that relaxes in
1\,nm in a nanowire or nanodot has a strain gradient 10$^3$ higher
than for a similar geometry in which a 1\% strain relaxes to zero
at the micron scale.  Thus, FxE can have a significant effect on the
properties of nanostructures. For example, a decrease in dielectric
constant\cite{catalan-jpcm04,catalan-prb05} and an increase in
critical thickness\cite{zhou-pb12} in thin films was attributed
to flexoelectric effects, and the transition
temperature and distribution of polarization can also be
significantly influenced.\cite{eliseev-prb09} A giant
enhancement of piezoelectric response\cite{majdoub-prb08} and
energy harvesting ability\cite{majdoub-prb08-1} were predicted in thin
beams. FxE was also shown to affect the properties of
superlattices\cite{aguado-puente-prb12}  and
domains walls.\cite{yudin-prb12,catalan-rmp12,borisevich-nc12} For example, 
domain configurations and polarization hysteresis curves 
were shown to be strongly affected by FxE because of
giant strain gradients present in epitaxial films.\cite{lee-prl11}
A FxE-induced rotation of polarization in certain domain walls
in \PTO\ was found,\cite{catalan-nm11} and purely mechanical writing of
domains in thin \BTO\ films was demonstrated.\cite{lu-s12}
Some piezoelectric devices based on the FxE effect have been proposed
and their effective piezoelectric response has been measured in
Cross's group.\cite{fousek-ml99,fu-jap06,chu-jap09}
Recently, it was found that a strain gradient can generate a
``flexoelectric diode effect''\cite{lee-nl12} based on a very
different principle from that of conventional diodes, such as $p-n$
junctions or Schottky barriers at metal-semiconductor interfaces,
which depend on the asymmetry of the system. The continuum theory
considering the FxE effect was also developed recently for the
nano-dielectrics\cite{shen-jmps10} and heterogeneous
membranes.~\cite{liu-pre13,mohammadi-jam13}

In order to understand the FxE response and to apply the FxE effect in
the design of functional devices, it is necessary to measure the FECs for
different materials.  Clearly it is desirable to look for materials
with large FECs, for direct applications of the FxE effect.  For
other nanoscale devices, on the other hand, it may actually be
desirable to identify materials where FxE is weak, so that unwanted
side effects of strain gradients are avoided.

Originally, FECs were estimated to be on the order of nC/m
and to scale linearly with static dielectric constant.\cite{kogan-spss64}
Forty years later, Ma and Cross found that the FECs, as measured by
beam bending experiments (see Sec.~\ref{sec:apxm}), could be three
orders of magnitude larger
than this in some high-$K$ materials.\cite{ma-apl01,ma-apl01-1,
ma-apl02,ma-apl03,ma-apl05,ma-apl06,cross-jms06}
This work set off a wave of related work
by other groups using a wide variety of approaches.
Using the same technique as Ma and Cross, Zubko
\textit{et al.}\cite{zubko-prl07,zubko-prl07-err} measured the FECs for
single-crystal \STO\ along different crystallographic
orientations in an attempt to
obtain the full FEC tensor. They report FECs for \STO\ that are on the
order of nC/m, much smaller than in previous experimental work.
They also found that it is impossible to obtain the full FEC
tensor through bending measurements alone, and that it is even difficult to
determine the sign of the effect. Another technique to measure FECs is to
apply uniaxial compression to a sample prepared in a truncated-pyramid
geometry, thus inducing a strain gradient in the pyramid.\cite{cross-jms06}
This also measures some kind of effective FEC, but does not easily
allow for extracting individual longitudinal components, due to
the complicated inhomogeneous strain gradient distribution.
However, based on this idea but using an inverse FxE effect, Fu
\textit{et al.}\cite{fu-jap06} measured
the FEC for Ba$_{0.67}$Sr$_{0.33}$TiO$_3$ and obtained the same
results as those from the direct FxE effect. Hana \textit{et al.}\ also used
the inverse FxE effect to measure the FEC for ceramic PMN-PT (a solid
solution of lead magnesium niobate and lead titanate).\cite{hana-f06,hana-f07}
By using a nanoindentation
method, Gharbi \textit{et al.}\cite{gharbi-ijss11} obtained the same order
of FEC for \BTO\ as in the work of Ma and Cross.\cite{ma-apl06}
Finally, Zhou \textit{et al.}\cite{zhou-epl12} proposed a method
to measure the flexocoupling coefficient by applying a homogeneous
electric field. They found that the hysteresis loop shifts due to
FxE effect, and that it can be restored by applying a 
homogeneous electric field. The size of the required electric field
was shown to be related to the
flexocoupling coefficient, and thus could be used to measure the FEC. This
method avoids the need to apply a dynamic mechanical load and may
increase the accuracy of measurement.

On the theoretical side, efforts to understand the FxE effect and
to extract FECs from theory began with the pioneering work of Kogan,
who first estimated FECs for simple dielectrics to be on the order
of nC/m and to scale linearly with static dielectric
constant.\cite{kogan-spss64} Twenty years later, Tagantsev developed
a model for FxE that was based on classical point-charge
models.\cite{tagantsev-prb86,tagantsev-pt91} The FxE response was
divided into four contributions denoted as ``static bulk,'' ``dynamic
bulk,'' ``surface FxE'' and ``surface piezoelectricity."
The first-principles electronic response was not accounted for
in this theory, which focused more on the lattice effects.
The calculation of static bulk FxE was later implemented by
Maranganti \textit{et al.},\cite{maranganti-prb09} and FECs for several
different materials were obtained. The FxE response of
two-dimensional systems were also investigated by Kalinin
\textit{et al.}\cite{kalinin-prb08} and Naumov
\textit{et al.}\cite{naumov-prl09}

The first attempt at a first-principles calculation of FECs for
bulk materials was carried out for \BTO\ and \STO\ by
Hong \textit{et al.},\cite{hong-jpcm10} who performed
calculations on supercells in which a longitudinal
strain variation of cosine form was imposed.  This gives
access to the longitudinal FEC $\mu_{1111}$, and implicitly
corresponds to fixed-$D$ (electric displacement field) electric
boundary conditions.  In this work, the positions of the Ba or Sr
atoms were fixed and other atoms were allowed to relax.
Their calculations include both electronic and lattice contributions
to the FECs, and their results show that FECs all take on
negative values. This method is limited to the longitudinal
contribution to the FEC tensor at fixed-$D$ boundary conditions.

Inspired by Martin's classical piezoelectric
theory,\cite{martin-prb72} Resta\cite{resta-prl10} developed
a first-principles theory of FxE, but it was
limited to the longitudinal electronic contribution to the FEC
response of elemental
materials, and was not implemented in
practice.  Shortly afterwards,
Hong and Vanderbilt\cite{hong-prb11} extended this theory to
general insulators and implemented it to calculate FECs for a variety
of materials, from elementary insulators to perovskites.
This theory was still limited to the electronic response, and only the
longitudinal $\m_{1111}$ components were computed in practice.
The calculations were at fixed-$D$ boundary conditions,
and several practical and convenient methods for achieving this
were proposed.  The results indicated that
the electronic FECs are all negative in sign and that they
do not vary very much between different materials classes.  This
work also raised several important issues, such as the dependence of
the results on choice of pseudopotential, the presence of surface
contributions related to the strain derivative of the surface
work function, and the need to introduce a current-density
formulation, instead of a charge-density one, to treat the transverse
components of the FEC tensor.

More recently, Ponomareva \textit{et al.}\cite{ponomareva-prb12} have
developed an approximate effective-Hamiltonian technique to study
FxE in \BSTO\ thin films in the paraelectric state at finite
temperature.  Parameters in the model are fit to first-principles
calculations on a small supercell in which an artificial periodic
strain gradient has been introduced.  The authors computed both the
flexocoupling coefficients (FCCs, see Sec.~\ref{sec:fcc}) and
FECs for \BSTO\ films for different thicknesses above the
ferroelectric transition temperatures. Unlike some of the previous
theories, they found all of the FEC tensor components to be positive.
They provided evidence that the dependence of the FEC tensor on thickness
and temperature basically tracked with the dielectric susceptibility,
suggesting a strategy in which FCCs are computed as a ``ground state
bulk property'' and the FECs scale with susceptibility. However,
since FEC calculations tend to be very sensitive to
the size of the supercell,\cite{hong-jpcm10} their small supercell
size may introduce a significant approximation.  They also did not
attempt to calculate the electronic contribution separately, and
the role of fixed-$\E$ vs.~fixed-$D$ electric boundary conditions
was not discussed.

In this manuscript, we present a complete first-principles theory
of flexoelectricity, based on a long-wave analysis of induced
dipoles, quadrupoles, and octupoles in the spirit of the
work of Martin\cite{martin-prb72} and Resta,\cite{resta-prl10}
together with an implementation via supercell calculations and
a presentation of computed values for a series of materials.
We find that the flexoelectric response can be divided into
``longitudinal'' and ``transverse'' components, and that the
treatment of the latter requires that one go beyond a
charge-response treatment to a current-response one.  The formalism
for this is presented in detail in Appendix~\ref{app:current}, but
its implementation is left for future work.  Therefore, we
present the longitudinal FEC tensor coefficients in their full
generality for our materials of interest, and in addition we present
some preliminary information about the transverse components.

During the preparation of this manuscript, we learned that M.~Stengel
has also developed a first-principles theory of
flexoelectricity\cite{stengel-unpub}
that bears many similarities
to the work presented here, although the point of view is different
in several respects.  Ref.~\onlinecite{stengel-unpub}
does not describe an actual implementation of the method in the
first-principles context or present any numerical results.
Nevertheless, the two theories seem to be in agreement on
fundamental points, and we hope that they will strengthen one
another.

In the remainder of this Introduction, we emphasize five points that
should be kept in mind when comparing calculated and/or measured
values of FECs.
Before proceeding, we refer the reader to several
useful review articles that have appeared recently covering both
experimental and theoretical aspects of the study of
flexoelectricity.\cite{tagantsev-pt91,cross-jms06,nguyen-am13,
zubko-armr13,yudin-unpub}

First, as indicated above, there are two contributions to the FECs:
a purely electronic (or ``frozen-ion'') contribution associated with
a naive set of atomic displacements that are simply quadratic
functions of their unperturbed positions, and
a lattice (or ``relaxed-ion'') contribution arising from additional
internal atomic displacements induced by the strain gradient.
Some of the previous theoretical work focused on the electronic
part,\cite{resta-prl10,hong-prb11} while others focused on the lattice
part,\cite{tagantsev-prb86,tagantsev-pt91,maranganti-prb09}, and still
others considered both implicitly but did not separate
them.\cite{hong-jpcm10,ponomareva-prb12}  In the present work,
we have developed a first-principles theory
which includes these two contributions explicitly, and we have proposed
a method to calculate them efficiently.

Second, the question of what, precisely, is meant by ``relaxed-ion''
is subtle for FxE and is discussed in Sec.~\ref{sec:relaxed-ion}
and Appendix~\ref{app:inverse}.  We find that the
calculation of the lattice contribution to the FECs depends on
a choice of ``force pattern'' applied to the atoms in the unit cell
in order to preserve the strain gradient, even after the induced
internal displacements have taken place.  It it possible to make
different choices for this force pattern. A mass-weighted
choice appears to be implicit in some previous work,
and is also appropriate to the analysis of dynamical long-wavelength
phonons.  However, other choices are possible, e.g., restricting the
forces to the A atoms of ABO$_3$ perovskites.\cite{hong-jpcm10}
We caution that
it is not meaningful to compare FECs computed using
different force patterns.  However, for an inhomogeneously strained
system in static equilibrium, the stress gradient
$\nabla \cdot \boldsymbol{\sigma}$ must vanish, where $\sigma$ is the
local stress tensor.  In such a case, the total FxE response is not
dependent on the choice of force pattern, since there is no
ambiguity about the meaning of relaxed atomic positions in this case.
This independence is confirmed by our numerical calculations.

Third, it is important to obtain all symmetry-independent elements
of the FEC tensor in order to understand the FxE response in the
case of an arbitrary strain distribution and to aid in the design
of functional FxE devices. However, it is challenging to obtain
the full FEC tensor for general materials, which have 54
independent components.\cite{quang-prsa11} Even for cubic
materials, which have only three independent components, there is
still no straightforward way to measure the full FEC tensor. Most
first-principles calculations have been limited to reporting the
longitudinal component in cubic materials,\cite{hong-jpcm10,hong-prb11}
although lattice (but not electronic) transverse components have
also been reported in some works.\cite{maranganti-prb09}
Here we develop a first-principles theory for the full FEC tensor.
However, our current implementation is still limited to longitudinal
components and to certain combinations of transverse components.
A formalism addressing the full set of transverse components is
presented in Appendix~\ref{app:current}, but the implementation of
such an approach is left to future work.
Moreover, we limit our formal considerations here to the case of
isotropic dielectric materials, since the convergence of some moment
expansions in Sec.~\ref{sec:charge-response} does not otherwise appear
to be guaranteed.

Fourth,
the reader should be aware that there are many different definitions
of FECs in the literature.  For example, the FECs can be defined in
terms of unsymmetrized strain, which tends to be more convenient
for the derivation of the formalism, or in terms of symmetrized
strain, which is more convenient in connecting to experimental
measurements.  A third object is the flexocoupling coefficient (FCC),
which appears directly in a Landau free-energy expansion.  Aside
from the confusion caused by the physical distinction between
these objects, there is also the practical problem that different
symbols and different subscript orderings are used for the same
quantity in different papers, making it very confusing when referring
to the FECs appearing in different contributions to the literature.
To help clarify this issue, we define the various kinds of FECs
(based unsymmetrized vs.~symmetrized strain) and the FCC, deriving
the transformations that can be used to convert between them.

Fifth, there is a well-know issue for FxE, namely the roughly
three-orders-of-magnitude discrepancy between most theoretical
estimates and experimental measurements of the FECs. Our work
suggests that this gap can largely be closed by paying close
attention to the difference between FECs computed at fixed $\E$
vs.~at fixed $D$.  This is important mainly for materials like
\BTO\ that have a large and strongly temperature-dependent
static dielectric constant $\epsilon^0$. The basic idea is to look for
quantities that scale only weakly with temperature, calculate
these from first principles, and then use the experimentally known
temperature dependence of $\epsilon^0$ to predict the FxE response
at elevated temperature.  Indeed, previous theory and calculations
predict that the FECs should scale linearly with dielectric
constant,\cite{kogan-spss64,tagantsev-prb86,ponomareva-prb12}
and experiments have also verified this.\cite{zubko-prl07}
Previous work has identified the FCC as an object with weak
temperature dependence that can be used in this way; as derived
from LGD theory,\cite{zubko-armr13,yudin-unpub} the FCC is roughly
the ratio between the FEC and $\epsilon^0$.  Indeed, Ponomareva
\textit{et al.}\cite{ponomareva-prb12} argue that the FCC
is ``ground-state bulk property'' that is independent of the
temperature and size of system. In the present work, we point out
that the FECs computed at fixed-$D$ boundary conditions are also
suitable for this purpose.  They are, in fact, closely related
to the FCCs, as we shall see in Sec.~\ref{sec:fcc}, but they
are more easily and directly computed from first principles.
Furthermore, we show formally in Sec.~\ref{sec:fixd} that the
ratio of the fixed-$\E$ to the fixed-$D$ FEC is just equal
to $\epsilon^0$.  Therefore, the strategy adopted here is to
compute the fixed-$D$ FECs at zero temperature and to use the known
temperature dependence of $\epsilon^0$ to make room-temperature
predictions.  Using this method, we find that the FECs computed
from our theory are much closer to the experimental values,
falling short by perhaps one order of magnitude instead of three.

The paper is organized as follows. In Sec.~\ref{sec:formalism}
we derive the charge-response formalism for the first-principles
theory of piezoelectricity and FxE, including electronic and
lattice contributions, based on the density of local dipoles,
quadrupoles and octupoles induced by a long-wave deformation.
Careful attention is paid to the lattice contributions, the
choice of fixed-$D$ and fixed-$\E$ electric boundary conditions,
and the form of the flexocoupling tensor and FxE tensor in
the case of cubic symmetry.  In Sec.~\ref{sec:calculation}
we propose a supercell approach for calculating the FEC tensor
for cubic materials based on the charge-response formalism.
Using two supercells, one extended along a Cartesian direction
and one rotated 45$^\circ$, we obtain all of the longitudinal
components of the response.  In Sec.~\ref{sec:results}, we report
the longitudinal FECs for different cubic materials at fixed $D$
from our first-principles calculations. The room temperature
FECs at fixed $\E$ are also obtained by using experimental
static dielectric constants. The full FEC tensor is also
reported after introducing some assumptions.
We then compare
our computed FECs with available experiment and theoretical
results. In Sec.~\ref{sec:conclusion} we give a summary and
conclusions. Finally, in the Appendices, we provide details about
three issues: the current-response formalism for full FEC tensors
(including longitudinal and transverse ones), the definition
and construction of the pseudo-inverse of force-constant matrix,
and the treatment of O atoms in the cubic perovskite structures.

\section{Formalism}
\mylabel{sec:formalism}
In 1972, Martin\cite{martin-prb72} introduced a theory of piezoelectricity
based on the density of local dipoles and quadrupoles induced by a
long-wave deformation (frozen acoustic phonon).  In recent years,
Martin's theory of piezoelectricity
has essentially been superseded by linear-response
and Berry-phase methods in the computational electronic structure
community.  These approaches only require consideration of a single
unit cell, and are therefore much more direct and efficient.  However,
they are derived from Bloch's theorem, so that while
they do apply to the case of a uniformly strained crystal, they do not
apply in the presence of a strain gradient.

To treat the problem of FxE, therefore, we follow Resta\cite{resta-prl10}
in returning to the long-wave method pioneered by
Martin.\cite{martin-prb72} For flexoelectricity, this requires an
analysis not only of induced dipoles and quadrupoles, but also of
induced octupoles.  While parts of our derivations are built upon
previous work,\cite{martin-prb72,tagantsev-prb86,tagantsev-pt91,
resta-prl10,hong-prb11,yudin-unpub} we attempt here to present a
comprehensive and self-contained derivation.

\subsection{General theory of charge-density response}
\mylabel{sec:charge-response}

We define
\beq
f\bit(\rr-\R_{lI})=\frac{\pa \rho (\rr)}{\pa u_{lI\t}}
\mylabel{eq:f}
\eeq
to be the change of charge density induced by the displacement
of atom $I$ in cell $l$, initially at $\R_{lI}$, by a distance
$u_{lI\t}$ along direction $\t$, keeping all other atoms fixed.
We also define the moments of the induced charge redistribution via
\bea
&&
Q\pone\biat=\int d\rr\, r\ba \,f\bit(\rr)\, ,
\mylabel{eq:Q1} \\
&&
Q\ptwo\biatb=\int d\rr\, r\ba\,  f\bit(\rr)\, r\bb \, ,
\mylabel{eq:Q2} \\
&&
Q\pthr\biatbc=\int d\rr\, r\ba\,f\bit(\rr)\, r\bb\, r\bc \,  .
\mylabel{eq:Q3}
\eea
Note that $Q\ptwo$ and $Q\pthr$ are symmetric under interchange
of $\a\b$ or $\a\b\c$ respectively.

It is worth briefly discussing the electric boundary conditions here.
For the displacement of a single atom, we do not have to specify
fixed $\E$ or $D$ boundary conditions; we just choose boundary
conditions such that the macroscopic potential and $\E$-field vanish
as $r^{-2}$ and $r^{-3}$ respectively.  Since the induced charge
density is screened, $Q\pone$ corresponds to the Callen dynamical
charge, not the Born one.  If instead of moving one atom we were
to move an entire sublattice, the Callen and Born charges would
correspond to the application of fixed-$D$ and fixed-$\E$ boundary
conditions respectively,  In this sense, we can regard the moment
tensors $Q\pone$, $Q\ptwo$ and $Q\pthr$ as being fixed-$D$ quantities.

The convergence properties of Eqs.~(\ref{eq:Q1}-\ref{eq:Q3}) also deserve
some discussion.  At large distances the induced charge
$f\bit(\rr)$ is mainly induced by the dipole field $\E\propto
r^{-3}$, raising questions about the convergence of the integrals
in Eqs.~(\ref{eq:Q1}-\ref{eq:Q3}) at large $r$.  Fortunately, 
$f\bit(\rr)$ also oscillates and averages to zero on the unit-cell
scale at large $r$, helping to tame any such divergences,
at least for the case of isotropic dielectrics (i.e.,
cubic materials).  This is less clear when the dielectric tensor
is anisotropic, so we accept Eqs.~(\ref{eq:Q1}-\ref{eq:Q3})
as being well-defined for isotropic dielectrics and focus on this
case in the rest of this paper.

We now introduce the {\it unsymmetrized} strain and strain gradient
tensors, defined as
\bea
&&
\e\bab = \frac{\pa u\ba} {\pa r\bb} \; ,
\mylabel{eq:unsymstrain}  \\
&&
\n\babc = \frac{\pa \e\bab} {\pa r\bc} = \frac{\pa^2 u\ba} {\pa
r\bb \pa r\bc} \;
\mylabel{eq:sgradient}
\eea
(here we think of $\rr$ as a spatial coordinate in the continuum
elasticity sense).
Note that $\e\bab$ is \textit{not} generally symmetric
($\e\bab \ne \e\bba$),
while $\n\babc$ is only symmetric in its last two indices
($\n\babc = \n_{\a\c\b}$).

We now consider a long-wavelength displacement wave (``frozen
acoustic phonon'') of wavevector $\k$ in the crystal,
so that the strain and strain gradient are given by
\beq
\u\pzer(\rr) = \u_{0} e^{i \k \cdot \rr},
\mylabel{eq:unom}
\eeq
the displacement $\u$, strain $\be$, and strain gradient $\bn$ are
\bea
&&
u\ba(\rr) = u_{0\a} \, e^{i \k \cdot \rr} \; ,
\mylabel{eq:displace} \\
&&
\e\bab(\rr) = i u_{0\a} k_{\b} \, e^{i \k \cdot \rr} \; ,
\mylabel{eq:strain} \\
&&
\n\babc(\rr) = - u_{0\a} k_{\b} k_{\c}\, e^{i \k \cdot \rr} \; .
\mylabel{eq:straing}
\eea
To a first approximation the atom displacements will follow the
nominal pattern of \equ{unom}, but the presence of the strain
and strain gradient may induce additional ``internal'' displacements
such that the the total displacement of atom $I$ in cell $l$
is\cite{martin-prb72}
\beq
\u_{lI} = (\u\pzer + \u\pone_I + \u\ptwo_I)\, e^{i \k \cdot \R_{lI}} \;.
\mylabel{eq:dis1}
\eeq
Here $\u\pzer$ is the ``acoustic'' displacement of the cell as a
whole (independent of atom index $I$),
$\u\pone_I$ is the additional displacement induced by strain $\be$, and
$\u\ptwo_I$ is the additional displacement induced by strain gradient $\bn$.
That is,
\bea
&&
u\pone \bit = \G \bitbc \, \e \bbc \; ,
\mylabel{eq:uone} \\
&&
u\ptwo \bit = N \bitbcd  \, \n \bbcd \; ,
\mylabel{eq:utwo}
\eea
where $\G\bitbc$ is the internal-strain tensor describing the
additional atomic displacements induced by a strain, and $N\bitbcd$
is the corresponding tensor describing the response to a strain
gradient.  Note that we adopt an implicit sum notation for
Greek indices representing Cartesian components of polarization,
field, or wavevector, such as $\b\c\d$ above, although we will
always write sums over the atomic displacement direction $\tau$
explicitly.

We shall see in Sec.~\ref{sec:relaxed-ion} how $\G\bitbc$
and $N\bitbcd$ can be related, via the force-constant matrix,
to force-response tensors $\L\bitbc$ and $T\bitbcd$ describing
the force $F\bit$ induced respectively by a strain or
strain gradient.  This is straightforward for $\L$, but we
shall see there that an ambiguity arises for $N$.
Deferring this issue for now, we substitute \equr{strain}{straing}
into \equr{uone}{utwo} to find
\bea
&&
u\pzer\bit
= u_{0\t} \, e^{i \k \cdot \R_{lI}} \; , \\
&&
u\pone\bit = i \G\bitbc \, u_{0\b} k\bc \, e^{i \k \cdot \R_{lI}} \; , \\
&&
u\ptwo\bit = - N\bitbcd \, u_{0\b} k\bc k\bd \, e^{i \k \cdot \R_{lI}}\; .
\eea
Defining
\beq
W_{I\t\b}(\k)=\delta_{\t\b} + i\G\bitbc \, k\bc - N\bitbcd \,
k\bc k\bd \; ,
\mylabel{eq:Wdef}
\eeq
Eq.~(\ref{eq:dis1}) can be written as
\beq
u_{lI\t} = W_{I\t\b}\, u_{0\b} \, e^{i \k \cdot \R_{lI}} \;.
\eeq
Then we can write down the induced charge density at $\rr$ as
\begin{widetext}
%
\beq
\rho(\rr)= \sum_{lI\t} f\bit(\rr - \R_{lI}) \, u_{lI,\t} =
\sum_{lI\t} f\bit(\rr - \R_{lI}) W_{I\t\b} \, u_{0\b} \, e^{i \k
\cdot \R_{lI}} \; ,
\mylabel{eq:rhof}
\eeq
and compute its Fourier transform as
\bea
\rho(\k) &=& V^{-1} \int {\rm d}\rr \,\rho(\rr) \, e^{-i \k \cdot \rr}  \nn
&=& \Vc^{-1} \sum_{I\tau} W_{I\t\b}(\k) \left(
\frac{1}{N}\sum_l \int d\rr\;f_{I\t}(\rr-\R_{lI}) e^{-i\k\cdot(\rr-\R_{lI})}
\right) u_{0\b} \nn
&=& \Vc^{-1} \sum_{I\tau} W_{I\t\b}(\k) \left(
\int d\rr'\;f_{I\t}(\rr') e^{-i\k\cdot\rr'} \right) u_{0\b} \;.
\mylabel{eq:rhokint}
\eea
(In going from the second to the third line above
we change the integration variable to $\rr'=\rr-\R_{lI}$,
notice that the result is independent of $l$, and cancel $\sum_l$ against
$N$, where $N$ is the number of cells of volume $\Vc$ in the total
system of volume $V$.\cite{explan-cell}) Thus we have
\beq
\rho(\k) = \Vc^{-1} \sum_{I\t} W_{I\t\b}(\k) \, f_{I\t}(\k) \,
u_{0\b} \; ,
\mylabel{eq:rhok}
\eeq
where the Fourier transform of $f_{I\t}(\rr)$ is
\bea
f_{I\t}(\k) &=& \int d\rr\;f_{I\t}(\rr) \, e^{-i\k\cdot\rr} \nn
&=& \int d\rr\;f_{I\t}(\rr) \left( 1 -ik_\mu r_\mu
-\frac{1}{2}k_\mu k_\nu r_\mu r_\nu
+\frac{1}{6}i\,k_\mu k_\nu k_\sigma r_\mu r_\nu r_\sigma +\ldots\right)
\;.
\mylabel{eq:fk-work}
\eea
Keeping terms up to third order in $\k$, we get
\beq
f_{I\t}(\k) =
-i \, k_\mu \, Q\pone_{I\mu\t}
-\frac{1}{2} \, k_\mu k_\nu \, Q\ptwo_{I\mu\t\nu}
+\frac{1}{6}i \, k_\mu k_\nu k_\sigma \, Q\pthr_{I\mu\t\nu\sigma}
\; ,
\mylabel{eq:fk}
\eeq
where the neutrality of the induced charge $f_{I\t}(\rr)$ has been
used to eliminate the zero-order term.

Plugging Eqs.~(\ref{eq:Wdef}) and (\ref{eq:fk}) into
Eq.~(\ref{eq:rhok}), we obtain an overall expansion of $\rho(\k)$
in powers of wavevector $\k$.  The linear-in-$k$ term vanishes by
the acoustic sum rule in the form $\sum_I Q\pone_{I\t\mu}=0$,
so we get
\bea
\rho(\k) &=& \Vc^{-1}\sum_I \left[
  \sum_\t Q\pone_{I\mu\t} \G_{I\t\b\c} \, k_\mu k_\c
  -\frac{1}{2}  Q\ptwo_{I\mu\b\nu} k_\mu k_\nu
\right] u_{0\b} \nn
&& +\Vc^{-1}\sum_I \left[
  i\sum_\t Q\pone_{I\mu\t} N_{I\t\b\c\d} \, k_\mu k_\c k_\d
  -\frac{1}{2} i \sum_\t Q\ptwo_{I\mu\t\nu} \G_{I\t\b\c} k_\mu k_\nu k_\c
  +\frac{1}{6} i Q\pthr_{I\mu\b\nu\sigma} k_\mu k_\nu k_\sigma
\right] u_{0\b} \nn
&& +\ldots
\mylabel{eq:expan}
\eea
%
\end{widetext}
%
We now want to relate this to the polarization $\P$, in terms
of which we can define the (unsymmetrized) piezoelectric
and flexoelectric tensors as
\beq
e_{\a\b\c} = \frac{\partial P_\a}{\partial\eta_{\b\c}}
\mylabel{eq:piezo-def}
\eeq
and
\beq
\mu_{\a\b\c\d} = \frac{\partial P_\a}{\partial\nu_{\b\c\d}}
\mylabel{eq:flexo-def}
\eeq
so that
\beq
P\ba = e_{\a\b\c}\, \e_{\b\c} + \m_{\a\b\c\d}\, \n_{\b\c\d} + \ldots
\mylabel{eq:Pexpan}
\eeq
Note that $\n_{\b\c\d}$ is symmetric in its last two indices $\c\d$,
so that $\m_{\a\b\c\d}$ is not uniquely specified by \equ{Pexpan};
to make it so, we adopt the convention
that $\m_{\a\b\c\d}$ is symmetric in $\c\d$ as well.
For the wave in question we find, using
Eqs.~(\ref{eq:strain}-\ref{eq:straing}),
\beq
P\ba(\k)=e_{\a\b\c}\,i\,u_{0\b}\,k_\c
+\mu_{\a\b\c\d}\,(-u_{0\b})\,k_\c k_\d + \ldots
\mylabel{eq:Peu}
\eeq
Using Poisson's equation in the form $\rho(\k) = -i k\ba
P\ba(\k)$, we obtain
\beq
\rho(\k)=e_{\a\b\c}\,k_\a k_\c \, u_{0\b}
+i\,\mu_{\a\b\c\d}\,k_\a k_\c k_\d \, u_{0\b} + \ldots
\mylabel{eq:expan2}
\eeq

Now, the strategy is to compare Eqs.~(\ref{eq:expan})
and (\ref{eq:expan2}) term-by-term in powers of $\k$.
Since the equation must be true for all $\u_0$ vectors for
a given $\k$, equating the second-order-in-$\k$ terms gives
%
\begin{widetext}
%
\beq
e_{\a\b\c}\,k_\a k_\c = \Vc^{-1}\sum_I \left[
\sum_{\t} Q\pone_{I\mu\t} \G_{I\t\b\c} \, k_\mu k_\c
  -\frac{1}{2}  Q\ptwo_{I\mu\b\nu} k_\mu k_\nu \right] \; ,
\mylabel{eq:safe-piezo}
\eeq
and similarly at the next order,
\beq
\mu_{\a\b\c\d}\,k_\a k_\c k_\d = \Vc^{-1}\sum_{I} \left[
\sum_{\t} Q\pone_{I\mu\t} N_{I\t\b\c\d} \, k_\mu k_\c k_\d
-\frac{1}{2} \sum_{\t} Q\ptwo_{I\mu\t\nu} \G_{I\t\b\c} k_\mu k_\nu k_\c
  +\frac{1}{6} Q\pthr_{I\mu\b\nu\sigma} k_\mu k_\nu k_\sigma
  \right]  \;.
\mylabel{eq:safe-flexo}
\eeq
These equations describe the piezoelectric and flexoelectric
responses respectively.

\subsection{Piezoelectric response}

We begin with the piezoelectric case.  From \equ{safe-piezo}
it follows that
\beq
e_{\a\b\c} =  \Vc^{-1} \sum_{I\t} Q\pone_{I\a\tau} \G_{I\tau\b\c} \,
 - \half \Vc^{-1} \sum_I Q\ptwo_{I\a\b\c} + A\babc
\;,
\mylabel{eq:e}
\eeq
\end{widetext}
where $A\babc$ is antisymmetric in the first and third indices
but otherwise arbitrary.  The first two
terms represent the lattice and electronic responses respectively.
The third vanishes under the symmetric sum over $\alpha\gamma$
on the left side of \equ{safe-piezo}, and serves as a reminder
that the forms given in the first two terms may not be fully
determined.  There is little danger of this regarding the first
term, which has a transparent interpretation in terms of
dipoles associated with strain-induced internal displacements of
the atomic coordinates.  Thus, we can write
\beq
e\babc = e\pld\babc + e\pel\babc \;,
\mylabel{eq:esum}
\eeq
where the lattice contribution
\beq
 e\pld\babc = \Vc^{-1} \sum_{I\t} Q\pone_{I\a\tau} \G_{I\tau\b\c}
 \;,
\mylabel{eq:eld}
\eeq
is denoted `ld' for `lattice dipole.'  The absence of a correction
term in \equ{eld} is demonstrated in Appendix~\ref{app:current}.
In the electronic term
$e\pel\babc$, however, a correction having the form of $A\babc$
cannot be discounted; relabeling $A$ as $e\pelT$, we obtain
\beq
 e\pel\babc = -\half\Vc^{-1} \sum_{I} Q\ptwo\biabc + e\pelT\babc
\;.
\mylabel{eq:eel}
\eeq
An explicit expression for $e\pelT\babc$ is given in
Appendix~\ref{app:current}.

We denote the first (symmetric in $\a\c$) and second (antisymmetric
in $\a\c$) terms of \equ{eel} as the ``longitudinal'' (L) and ``transverse''
(T) parts respectively.
To clarify this terminology,
note that a given deformation
of the medium will generate a polarization field $\P(\rr)$ whose
longitudinal and transverse parts are defined as the curl-free
and divergence-free portions respectively, so that any
piezoelectrically-induced charge density $\rho=-\nabla\cdot\P$
comes only from the longitudinal part.  But a simple calculation
shows that
\bea
\partial_\a P\ba(\rr) &=& \partial\ba(e\babc\,\eta\bbc) \nn
        &=& e\babc\,\n\bbac \nn
        &=& e^\textrm{S}\babc\,\n\bbac \;,
\mylabel{eq:gradpp}
\eea
where the symmetry of $\n\bbac$ under $\a\c$ is
used in the last step.
This shows that the ``symmetric'' and ``antisymmetric''
parts of $e\babc$ are indeed just the longitudinal and transverse
contributions, respectively.

Recall that this is the piezoelectric response to the unsymmetrized
strain tensor of \equ{unsymstrain}, and so contains responses
to the rotation of the medium as well as to a symmetric strain.
Defining the symmetric and antisymmetric parts as
\bea
\str\bab &=& (\eta\bab+\eta\bba)/2 \;, \mylabel{eq:symstr}\\
\omega\bab &=& (\eta\bab-\eta\bba)/2 \;,
\eea
and considering the general case
$e\babc=e^{\textrm{S}}\babc+e^{\textrm{A}}\babc$ (with $e^{\textrm{S}}$
and $e^{\textrm{A}}$ respectively symmetric and antisymmetric under
indices $\a\c$), one finds that
\beq
P\ba=e^\textrm{S}\babc\,\str\bbc+e^{\textrm{A}}\babc\,\omega\bbc
\;.
\eeq
Here the antisymmetric part corresponds to the change of
polarization resulting from rotation of the crystal,
and thus contributes to the ``improper'' piezoelectric response.
\cite{vanderbilt-jpcs00}  However, the improper response also
includes symmetric contributions (e.g., from volume-nonconserving
symmetric strains), so the ``proper'' piezoelectric tensor cannot
simply be equated with $e^{\textrm{S}}\babc$.
After a careful analysis that made use of sum rules associated
with uniform translations and rotations of the lattice, Martin
\cite{martin-prb72} was able to show that the proper piezoelectric
response is given by Eqs.~(\ref{eq:esum}) and (\ref{eq:eld})
with \equ{eel} replaced by
\beq
e^{\textrm{el,prop}}\babc = -\half \Vc^{-1} \sum_{I} \left[
 Q\ptwo_{I\a\b\c} - Q\ptwo_{I\c\a\b} + Q\ptwo_{I\b\c\a}
\right] \;.
\mylabel{eq:eel2}
\eeq
Thus, while it is far from obvious, it turns out that the proper
piezoelectric tensor depends only on the symmetric parts of
the response.  This is consistent with simple counting arguments:
the tensor describing the polarization response to a symmetric
strain has 18 independent elements, as does $Q\ptwo_{I\a\b\c}$.

Interestingly, the distinction between proper and improper
responses does not arise for flexoelectricity, which is defined in
terms of the polarization at a point in the material at which
$\eta\bab$ is zero (although the strain gradient is not).
Also,
while the symmetrized strain tensor
contains less information than the unsymmetrized $\eta\bab$ (six elements
vs.~nine), this is not true of strain gradients.  Instead, the
symmetrized and unsymmetrized strain gradients contain the
same information
(18 unique elements) and are related by\cite{zubko-thesis}
\beq
\n\bbcd = \frac{\pa^2 u\bb}{\pa r\bc \pa r\bd} = \frac{\pa
\str_{\b\c}} {\pa r\bd} + \frac{\pa \str_{\b\d}}{\pa r\bc} -
\frac{\pa \str_{\c\d}}{\pa r\bb} \;.
\mylabel{eq:strgrad}
\eeq
In this one respect, the treatment of the flexoelectric response is
actually simpler than for the piezoelectric one.

\subsection{Flexoelectric response}
\label{sec:flexo-part}

We turn now to the flexoelectric response.
The general solution of \equ{safe-flexo} is
%
\begin{widetext}
%
\beq
\m\babcd =  \Vc^{-1} \sum_{I\t}
Q\pone_{I\a\tau} N_{I\tau\b\c\d} \,
- \frac{1}{4}\Vc^{-1} \sum_{I\t}
 \left( Q\ptwo_{I\a\tau\d} \G_{I\tau\b\c}
      + Q\ptwo_{I\a\tau\c} \G_{I\tau\b\d} \right)
 + \frac{1}{6} \Vc^{-1} \sum_I Q\pthr_{I\a\b\c\d}
 + B\babcd \;,
\mylabel{eq:flexo-big}
\eeq
%
\end{widetext}
%
where the $Q\ptwo$ term has been symmetrized to obey the
requirement that $\mu\babcd$ be symmetric in $\c\d$, and
$B\babcd$ is an extra ``antisymmetric'' piece.
For our purposes we define the ``symmetric part'' of $X\babcd$
(that is symmetric in its last two indices) to be
\beq
X^\textrm{S}\babcd=\frac{1}{3}(X_{\a\b\c\d}+ X_{\c\b\d\a}+ X_{\d\b\a\c})
\mylabel{eq:symmdef}
\eeq
and the antisymmetric part to be $X^\textrm{A}=X-X^\textrm{S}$.
So, we are allowed to add an extra antisymmetric term
$B=B^\textrm{A}$ to \equ{flexo-big} because 
it will vanish under the sum over $\a\c\d$ in \equ{safe-flexo}.

Clearly \equ{flexo-big} contains
three terms, two of which involve lattice responses.  We write
\beq
\mu\babcd=\mu\pld\babcd+\mu\plq\babcd+\mu\pel\babcd \;,
\mylabel{eq:mu-three}
\eeq
where the terms on the right side are the lattice dipole, lattice
quadrupole, and electronic terms, respectively.  Writing these
explicitly,
\bea
&&\mu\babcd\pld=\Vc^{-1} \sum_{I\t} Q\pone_{I\a\tau} N_{I\tau\b\c\d} \;,
\mylabel{eq:muld} \\
&&\mu\babcd\plq=-\frac{1}{4}\Vc^{-1} \sum_{I\t} \left(
    Q\ptwo_{I\a\tau\d} \G_{I\tau\b\c} \right. \nn
  && \hspace{2.6cm}  \left. + Q\ptwo_{I\a\tau\c} \G_{I\tau\b\d} \right)
      +\mu\babcd\plqJ \;,
\mylabel{eq:mulq} \\
&&\mu\babcd\pel=\frac{1}{6} \Vc^{-1} \sum_I Q\pthr_{I\a\b\c\d}+\mu\babcd\pelJ
\;.
\hspace{1.6cm}
\mylabel{eq:muel}
\eea
where the last terms in \equs{mulq}{muel} are extra antisymmetric
contributions and $\mu\plqJ+\mu\pelJ$ corresponds to the $B$
term in \equ{flexo-big}.  The label ``J'' indicates that these
terms arise from the current-response formulation given in
Appendix~\ref{app:current}; explicit expressions for these
corrections, and a demonstration that no correction is needed
for $\mu\pld$, are given there.

Let us emphasize again the physics of these corrections.
First, we can straightforwardly extend the discussion
at the end of the last subsection to the case of flexoelectricity.
In place of \equ{gradpp} we find
\bea
\partial_\a P\ba(\rr) &=& \partial_\a(\m\babcd\,\nu\bbcd) \nn
        &=& \m\babcd\,h_{\b\a\c\d} \nn
        &=& \m^\textrm{S}\babcd\,h_{\b\a\c\d}
\mylabel{eq:gradpf}
\eea
where
\beq
 h_{\b\a\c\d} = \frac{\pa\e\bbcd}{\pa r\ba}
=\frac{\pa^3 u\bb}{\pa r\ba \pa r\bc \pa r\bd}
\mylabel{eq:sigdef}
\eeq
is fully symmetric in the last three indices $\a\c\d$.
It again follows that ``symmetric'' and ``antisymmetric''
correspond to ``longitudinal'' (L) and ``transverse'' (T)
respectively.

Now the essential problem is that the
charge density appearing in \equ{rhof}, used as the starting point
of the derivation given above, is only sensitive to the longitudinal
response, since it only depends on the divergence of $\P(\rr)$.
Thus, the expression given for the FEC tensor in \equ{flexo-big},
excluding the final antisymmetric $B\babcd$ term, must contain
\textit{all} of the longitudinal response, but may contain only
part of, or may omit altogether, the transverse response.
A simple calculation shows that the $Q\pone\,N$ and $Q\ptwo\,\G$
terms in \equs{muld}{mulq} do contain transverse parts,
while the $Q\pthr$ term in \equ{muel} does not.
The last terms in \equs{mulq}{muel}
are contributions to the transverse parts $\mu\plqT$ and
$\mu\pelT$ of the lattice-quadrupole and electronic
responses.

While the transverse parts $\mu\plqT$ and $\mu\pelT$ make no
contribution to the induced internal charge density $\rho(\rr)$,
this does not mean that the transverse terms have no
physical consequence.  Polarization-related bound charges also
arise at surfaces and interfaces of the sample, and these can
depend on the transverse as well as the longitudinal part of the
flexoelectric response, as occurs for example for beam-bending
geometries as discussed in Sec.~\ref{sec:apxm}.  Thus, a full
theory of flexoelectricity should contain both contributions,
as derived in Appendix~\ref{app:current}.
In the remainder of this manuscript, however, we concentrate on
computing the longitudinal contributions alone.

Finally, we note that the need for transverse corrections is also
evident from counting arguments. For example, looking at the
electronic contribution of \equ{muel}, we can see that $\mu\babcd\pel$
has 54 independent tensor elements (3$\times$3$\times$6 since
it is symmetric under $\c\d$) while $Q\pthr_{I\a\b\c\d}$ has
only 30 (3$\times$10 since it is symmetric under $\a\c\d$).
Thus, the $Q$ moment tensors do not contain enough information
to fully specify the flexoelectric response.  On the other hand,
the symmetric (i.e., longitudinal) part of $\m\pel$ has only 30
independent elements and can thus be captured by $Q\pthr_{I\a\b\c\d}$.

\subsection{Crystals with cubic symmetry}

For crystals with full cubic point symmetry, the piezoelectric tensor
vanishes by symmetry and the flexoelectric tensor
$\mu\babcd$ has only three independent elements,\cite{quang-prsa11}
namely $\m\bl$, $\m\btm$, and $\m\btr$.  Others related by
interchange or cycling of Cartesian indices are equal
(e.g., $\m_{1221}= \m_{3113}$) and those with any Cartesian
index appearing an odd number of times (e.g, $\m_{1223}$) vanish.

Using these relations and \equ{sigdef} we can explicitly write
$-\rho(\rr)=A\,\m\bl+B\,\m\btr+C\,\m\btm$
with $A=h_{1111}+h_{2222}+h_{3333}$,
$B=h_{1122}+h_{2211}+h_{1133}+h_{3311}+h_{2233}+h_{3322}$,
and $C=2B$.  The internal bound charge resulting from the
flexoelectric response to the deformation is then proportional
to $A\,\m\bl+B\,(\m\btr+2\m\btm)$.  This motivates us to define
a new set of three coefficients as
\bea
&&\mlone = \mu\bl  \; ,
\mylabel{eq:mlone} \\
&&\mltwo = \mu\btr + 2 \mu\btm \; ,
\mylabel{eq:mltwo} \\
&&\mtr =  \mu\btr - \mu\btm \;.
\mylabel{eq:mtr}
\eea
Here `L1' and `L2' indicate ``longitudinal'' terms which
contribute to the internal bound charges in proportion
to combinations $A$ and $B$ respectively, while
`T' indicates a ``transverse'' term.
Thus, we see that a general cubic material is characterized
by two longitudinal and one transverse flexoelectric
coefficient.
For a material such as glass that has isotropic symmetry,
one finds that $\m\bl = \m\btr + 2 \m\btm$,
i.e., $\mlone=\mltwo$, in which case there is only one
longitudinal coefficient.  Thus, we can think of
$\Delta=\mltwo-\mlone$ as a measure of the anisotropy of the cubic
medium, which shows up only in the longitudinal response.

In general, the flexoelectric response of a cubic crystal
can have contributions from all three terms in \equ{mu-three}.
However, as we shall see in Sec.~\ref{sec:modes},
the lattice quadrupole
term of \equ{mulq} vanishes in simple cubic materials including
those found in rocksalt, cesium chloride, and perovskite crystal
structures.  This term can be non-zero in more complex cubic
materials, such as spinels and pyrochlores; the technical
requirement is the presence of zone-center Raman-active phonon
modes, or equivalently, the existence of free Wyckoff parameters.
This will be discussed further in Sec.~\ref{sec:modes}.

\subsection{Definitions in terms of symmetrized strains}
\mylabel{sec:symstr}

There are many different definitions of FECs in the literature.
Up until now we have been working with the
{\em unsymmetrized} strain tensor
$\eta\bab=\partial u\ba/\partial r\bb$
and its gradient $\nu\babc =\partial\eta\bab/\partial r\bc$ defined
in Eqs.~(\ref{eq:unsymstrain}-\ref{eq:sgradient}); this form
is convenient for formal derivations and for practical
calculations, and corresponds to the $\mu$ in 
Refs.~[\onlinecite{hong-prb11,maranganti-prb06,maranganti-prb09}]
and the $f$ in Ref.~[\onlinecite{tagantsev-pt91}].
On the other hand, the
FEC related to {\em symmetrized} strain is convenient for
experimental measurements; see, e.g., $g$ in
Ref.~[\onlinecite{zubko-thesis}], $f$ in Refs.~[\onlinecite{catalan-nm11,
gharbi-apl09,hong-jpcm10,lee-prl11,lu-s12,zubko-prl07}],
$\mu$ in Refs.~[\onlinecite{ma-apl01,ma-apl01-1,ma-apl02,ma-apl03,
ma-apl05,ma-apl06,chu-jap09,cross-jms06,fu-jap06,gharbi-ijss11,
majdoub-prb09-2,ponomareva-prb12,resta-prl10,sharma-jmps07}],
$F$ in Ref.~[\onlinecite{quang-prsa11}], and $\gamma$ in
Ref.~[\onlinecite{marvan-pcps88}].
Researchers sometimes use different definitions without emphasizing
their relations.  Complicating matters further is the fact that
different conventions are frequently used in the literature for the
order of the four subscript indexes of the FEC tensor,
both for unsymmetrized and symmetrized strain cases,
which can cause confusion especially for the transverse
components.  In this section, therefore, we clarify the relations
between the unsymmetrized and symmetrized formulations following
the analysis in P.~Zubko's thesis.\cite{zubko-thesis} Throughout
our paper, we use the notation $\mu$ and $g$ for the FECs defined
in terms of unsymmetrized and symmetrized strains respectively.

We define the gradient of the symmetrized strain as
\beq
\nb\bbcd=\frac{\partial\str\bbc}{\partial r\bd}
=\frac{1}{2}(\n\bbcd+\n_{\c\b\d})
\;.
\eeq
Note that $\nb\bbcd$ is symmetric in the first two indices $\b\c$,
while instead $\n\bbcd$ is symmetric in the last two indices
$\c\d$.  The inverse relation to the above equation is
\beq
\n\bbcd = \frac{\pa^2 u\bb}{\pa r\bc \pa r\bd} =
\nb_{\b\c\d}+\nb_{\b\d\c}-\nb_{\c\d\b} \;,
\mylabel{eq:nu-nus}
\eeq
which appeared earlier as \equ{strgrad}.  In the context of
symmetrized strains, we then define the flexoelectric coefficient
$\gtil\babcd$ to obey (note the order of indices)
\beq
P\ba = \gtil_{\a\d\b\c} \, \nb\bbcd \;.
\mylabel{eq:gtil}
\eeq
Comparing this with
\beq
P\ba = \mu\babcd \, \n\bbcd \; ,
\eeq
it follows that
\beq
\gtil_{\a\d\b\c} = \m\babcd + \m_{\a\b\d\c} - \m_{\a\d\b\c} \; .
\mylabel{eq:g-m}
\eeq

Recall that we defined $\m\babcd$ to be symmetric in $\c\d$
by convention.  Then $\gtil_{\a\d\b\c}$ as given by \equ{g-m}
is \textit{not} generally symmetric in its own last indices
$\b\c$.\cite{zubko-thesis}  Alternatively, we may define
\bea
g_{\a\d\b\c} &=& \half (\gtil_{\a\d\b\c} + \gtil_{\a\d\c\b}) \nn
 &=& \mu_{\a\b\c\d} + \mu_{\a\c\b\d} - \mu_{\a\d\b\c} \; .
 \mylabel{eq:sym-FEC}
\eea
This \textit{is} symmetric in $\b\c$, making it a
more natural definition in the symmetrized-strain
context, where also $\nb\bbcd$ is symmetric in $\b\c$.
In this case, however, the $\tilde\m\babcd$ that is
related to $g_{\a\d\b\c}$ by the analog of
\equ{g-m} is no longer symmetric in its own last indices
$\c\d$.  By convention, $\mu$ and $g$ are
usually used in the unsymmetrized-strain and symmetrized
strain contexts respectively, so \equ{sym-FEC} should
be used to do the conversion instead of \equ{g-m}. ~\cite{notes-mu-g}

For a cubic system we have
\bea
&&g_{1111} = \m_{1111} \;, \mylabel{eq:gma}\\
&&g_{1122} = 2\m_{1221}-\m_{1122} \;, \mylabel{eq:gmb}\\
&&g_{1221} = \m_{1122} \;,    \mylabel{eq:gmc}
\eea
and the flexoelectric coefficients defined in
Eqs.~(\ref{eq:mlone}-\ref{eq:mtr}) can be written as
\bea
&&\mlone =g\bl  \; , \\
&&\mltwo =g\btr + 2g\btm \; , \\
&&\mtr = -\half(g\btr -g\btm) \; .
\eea

\subsection{Lattice contributions}
\mylabel{sec:relaxed-ion}

We return now to the lattice (or ``relaxed-ion'') contributions
to the flexoelectric response, given by \equs{muld}{mulq},
neglecting now the current-response contribution to the latter.
In Sec.~\ref{sec:charge-response} we defined
\bea
&& \G\bitbc=\frac{\pa u\pone\bit}{\pa\e\bbc} \;,
\mylabel{eq:Gdef}\\
&&  N\bitbcd=\frac{\pa u\ptwo\bit}{\pa\n\bbcd} \;,
\mylabel{eq:Ndef}
\eea
which are the ``internal-strain'' tensors describing the
\textit{displacements} of the atoms in response to a strain or
strain gradient respectively.  Correspondingly, we define
\bea
&& \L\bitbc=\frac{\pa F\bit}{\pa\e\bbc} \;,
\mylabel{eq:Ldef}\\
&&  T\bitbcd=\frac{\pa F\bit}{\pa\n\bbcd} \;,
\mylabel{eq:Tdef}
\eea
representing the \textit{forces} appearing on the atoms due
to a homogeneous strain or strain gradient.

For the strain-induced case we assume that the atoms adjust to their
equilibrium positions as the strain is applied.  The force balance
equations then take the form $0=dF\bit/d\e\bbc$, or
\bea
0 &=& \frac{\pa F\bit}{\pa \e\bbc}
 +\sum\bjtp \frac{\pa F\bit}{\pa u\pone\bjtp}
         \, \frac{\pa u\pone\bjtp}{\pa\e\bbc} \nn
&=& \L\bitbc - \sum\bjtp K_{I\t,J\t'} \, \G\bjtpbc \;,
\mylabel{eq:fbal-str}
\eea
where
\beq
K_{I\t,J\t'}=-\frac{\pa F\bjtp}{\pa u\bit} \;,
\mylabel{eq:Kdef}
\eeq
is the zone-center force-constant matrix.  It follows that
\beq
\G\bitbc=\sum\bjtp \Ki_{I\t,J\t'}\,\L\bjtpbc \;,
\mylabel{eq:GKL}
\eeq
where $\Ki$ is the pseudo-inverse of $K$
(see Sec.~\ref{sec:force-pattern}).  When inserted in
\equ{eld}, this gives the standard result for the lattice
piezoelectric response,
\beq
 e\pld\babc = \Vc^{-1} \sum_{I\t,J\t'} Q\pone_{I\a\tau}
   \Ki_{I\t,J\t'} \, \L\bjtpbc \;,
\mylabel{eq:eldz}
\eeq
and a similar substitution can be made in \equ{mulq} for the
lattice-quadrupole flexoelectric response to get
\beq
\mu\babcd\plq=-\frac{\Vc^{-1}}{4} \sum_{I\t,J\t'}
    Q\ptwo_{I\a\tau\d} \,\Ki_{I\t,J\t'} \, \L\bjtpbc + \ldots
\mylabel{eq:mulqz}
\eeq
where the `$\ldots$' refers to the term with $(\c,\d)$ interchanged.

For the lattice-dipole flexoelectric response of \equ{muld}, we
would similarly like to write the force-balance equations
\bea
0 &=& \frac{\pa F\bit}{\pa \n\bbcd}
 +\sum\bjtp \frac{\pa F\bit}{\pa u\ptwo\bjtp}
         \, \frac{\pa u\ptwo\bjtp}{\pa\n\bbcd} \nn
&=&  T\bitbc - \sum\bjtp K_{I\t,J\t'} \,  N\bjtpbc \;,
\mylabel{eq:fbal-gra}
\eea
which would lead to
\beq
 N\bitbcd=\sum\bjtp \Ki_{I\t,J\t'}\, T\bjtpbcd \;,
\mylabel{eq:NKT}
\eeq
so that \equ{muld} becomes
\beq
\m\babcd\pld=\Vc^{-1} \sum_{I\t,J\t'} Q\pone_{I\a\tau}
  \,\Ki_{I\t,J\t'}\,T_{J\t'\b\c\d} \;.
\mylabel{eq:muldT} \\
\eeq
This is how we calculate lattice flexoelectric response in
this work; we first compute $Q\pone$, $K$ and $T$
from our first-principles calculations, and then combine them
via \equ{muldT}.

Strictly speaking, however,
\equ{fbal-gra} has no solution, for the simple reason that a
true force balance is not possible: relaxing the atoms to their
equilibrium positions would erase the strain gradient.
Formally, the problem is that when summed over the atom index
$I$, the first term of \equ{fbal-gra} is generally non-zero, while
the second vanishes by the acoustic sum rule.  Physically, the
problem is that a strain gradient is always accompanied by a
stress gradient, which in general gives rise to a force density.
Thus, external forces, not accounted for in \equ{fbal-gra},
need to be applied to the atoms in each unit cell in order
to oppose this force density.  As will be discussed in
Sec.~\ref{sec:force-pattern}, we can still use \equ{NKT} as
long as $K^{-1}$ is replaced by an appropriately chosen
pseudo-inverse.  There is some freedom in the choice of this
pseudo-inverse, but physical results for static deformations,
such as the beam-bending configurations discussed in
Sec.~\ref{sec:apxm}, will ultimately be independent of this choice.

Finally, following Tagantsev,\cite{tagantsev-prb86}
we note that the zone-center force-constant matrix
$K_{I'\t,J\t'}$ and the
force-response tensors $\L\bitbc$ and
$T\bitbcd$ can themselves be written in a manner
somewhat parallel to \equr{Q1}{Q3}, but this time as moments
of the full force-constant matrix
\beq
\Phi^{lIJ}_{\t\t'}=-\frac{\pa F_{0I\t}}{\pa u_{lJ\t'}} \;.
\mylabel{eq:Phi}
\eeq
Using
\bea
\L\bitbc &=& \frac{\pa F_{0I\t}}{\pa \e_{\b\c}}
    = \sum_{lJ\t'}\frac{\pa F_{0I\t}} {\pa u\pzer_{lJ\t'}}
       \, \frac{\pa u\pzer_{lJ\t'}}{\pa \e_{\b\c}} \;,\\
T\bitbcd &=& \frac{\pa F_{0I\t}}{\pa \n_{\b\c\d}}
    = \sum_{lJ\t'}\frac{\pa F_{0I\t}} {\pa u\pzer_{lJ\t'}}
       \, \frac{\pa u\pzer_{lJ\t'}}{\pa \n_{\b\c\d}} \;,
\mylabel{eq:TFn}
\eea
it follows that
\bea
K_{I\t,J\t'} &=& \sum_l \Phi^{lIJ}_{\t\t'}  \;,\\
\L\bitbc &=& - \sum_{lJ}
\Phi^{lIJ}_{\t\b} \, \Delta R^{lIJ}_{\c} \;,\\
T\bitbcd &=& -\half \sum_{lJ}
\Phi^{lIJ}_{\t\b} \, \Delta R^{lIJ}_{\c} \, \Delta R^{lIJ}_{\d} \;,
\mylabel{eq:T1}
\eea
where $\Delta R^{lIJ}_\b=(\R_{lJ}-\R_{0I})_\b$.
As a reminder, $ll'$ are unit cell labels while $IJ$ and
$\t\t'$ are atom and displacement-direction labels respectively.
Regarding the convergence of these sums at large
$\Delta R$, considerations similar to those discussed
below Eqs.~(\ref{eq:Q1}-\ref{eq:Q3}) apply.
The practical calculation of the elements of the $T$ tensor
will be described in Sec.~\ref{sec:orig-frame}, where
\equ{T1} takes the form of \equ{Tsup} after being adapted to
the supercell context.

\subsubsection{Transformation to mode variables}
\mylabel{sec:modes}

In simple binary crystals such as NaCl or CsCl, the above formulas
can be used directly.  For more complicated crystals such as
perovskites, however, it is useful to carry out a transformation
to symmetry-mode variables.  Here we briefly sketch the
transformation to an arbitrary set of mode variables, and then
discuss in particular the case of symmetry modes chosen according
to the irreducible representations (irreps) of the zone-center
force-constant matrix.

Let $\xi_j$ ($j=1,...,3N$) be a set of mode variables that are
related to the $3N$ atomic displacements according to
\beq
\xi_j=\sum\bit A\bjit\,u\bit \;,
\eeq
with $A\bjit$ expressing the linear transformation from one
basis to the other.  The inverse relation is
\beq
u\bit=\sum_j\Ai\bitj\,\xi_j \;.
\eeq
Then using a tilde to indicate quantities expressed in the mode
representation, the various quantities of interest transform as
\bea
\Qt\pone_{j\a} &=& \sum\bit \Ai\bitj \,Q\pone\biat \;, \\
\Qt\ptwo_{j\a\b} &=& \sum\bit \Ai\bitj \,Q\ptwo\biatb \;, 
\mylabel{eq:aiq} \\
\Kt_{ij} &=& \sum_{IJ\t\t'} \Ai_{I\t,i}\,\Ai_{J\t',j}\,
   K_{I\t,J\t'} \;, \\
\Lt_{j\b\c} &=& \sum\bit \Ai\bitj \, \L\bitbc\;, \\
\Tt_{j\b\c\d} &=& \sum\bit \Ai\bitj \,  T\bitbcd\;.
\mylabel{eq:Ttrans}
\eea
Then Eqs.~(\ref{eq:eld}), (\ref{eq:muld}), and (\ref{eq:mulq})
become
\bea
 e\pld\babc &=& \Vc^{-1} \sum_j \Qt\pone_{j\a} \, \Gt_{j\b\c} \;, \\
 \m\pld\babcd &=& \Vc^{-1} \sum_j \Qt\pone_{j\a} \,  \Nt_{j\b\c\d} \;, \\
 \m\plq\babcd &=& -\frac{1}{4}\Vc^{-1} \sum_j \left(
      \Qt\ptwo_{j\a\d} \, \Gt_{j\b\c} \right.
\mylabel{eq:mulqt} \nn
&& \hspace{1.7cm} \left.
     +\Qt\ptwo_{j\a\c} \, \Gt_{j\b\d} \right) \;,
\eea
where \equs{GKL}{NKT} have been replaced by
\bea
\Gt_{i\b\c} &=&\sum_j(\Kt^{-1})_{ij}\,\Lt_{j\b\c} \;, \\
\Nt_{i\b\c\d} &=&\sum_j(\Kt^{-1})_{ij}\,\Tt_{j\b\c\d} \;.
\eea
Note that the $\m\plqJ$ term of \equ{mulq}
has been omitted in \equ{mulqt} above, but can
easily be restored by converting $J\poneT\biabc$ of
\equ{mulqj} into the mode representation in a manner analogous
to \equ{aiq}.

This formulation becomes especially advantageous if the mode
variables are chosen to be symmetry-adapted.  Let us relabel the
modes as $j\rightarrow\{s\sigma a\}$, where $s$ is the irrep
label, $\sigma$ labels the copy of the irrep if there is more than
one, and $a=1\ldots m_s$ ($m_s$ is the dimension of
irrep $s$) labels the basis vectors.
Then the zone-center force-constant
matrix $K$ is diagonal in $s$ and $a$,
\beq
\Kt_{s\sigma a,s'\sigma'a'}=\delta_{ss'}\,\delta_{aa'}\,
     \tilde{k}^s_{\sigma\sigma'} \;,
\eeq
and its pseudo-inverse can be written similarly but using
$\kti_s$ which is the $m_s\times m_s$ pseudo-inverse
of $\tilde{k}^s$.  In this notation we have
\bea
 e\pld\babc &=& \Vc^{-1} \sum_{s\sigma\sigma'a} \Qt\pone_{s\sigma a,\a}
   \, \kti_{s,\sigma\sigma'} \, \Lt_{s\sigma'a,\b\c} \;,
\mylabel{eq:eldsy} \\
 \m\pld\babcd &=& \Vc^{-1} \sum_{s\sigma\sigma'a} \Qt\pone_{s\sigma a,\a}
    \, \kti_{s,\sigma\sigma'}\, \Tt_{s\sigma'a,\b\c\d} \;,
\mylabel{eq:muldsy} \\
 \m\plq\babcd &=& -\frac{1}{4}\Vc^{-1} \sum_{s\sigma\sigma'a} \left(
      \Qt\ptwo_{s\sigma a,\a\d} \,\kti_{s,\sigma\sigma'} \,
    \Lt_{s\sigma'a,\b\c} \right. \nn
&& \hspace{1.3cm} \left.
     +\Qt\ptwo_{s\sigma a,\a\c} \, \kti_{s,\sigma\sigma'} \,
   \Lt_{s\sigma'a,\b\d} \right) \;.
\mylabel{eq:mulqsy}
\eea

It is clear that $\Qt\pone$, which describes the
electric dipole appearing in response to a mode displacement,
behaves like a Cartesian vector,
and so will only be nonzero for vector irreps.
Thus the sums over $s$ in \equs{eldsy}{muldsy} can
be restricted to irreps of vector character, and $\Lt$ and
$\Tt$ need only be evaluated for these, i.e., for infrared-active
modes.  For cubic materials there is just one such irrep,
namely $T_1^-$ (also known as $\Gamma_{15}^-$).
It is well-known for piezoelectricity that only infrared-active
modes contribute, but the above makes it clear that the same is
true for the lattice-dipole flexoelectric response.

As for the lattice-quadrupole contribution in \equ{mulqsy},
$\Qt\ptwo$ describes the electric quadrupole response, which
has the character of a symmetric
second-rank tensor and thus only couples to quadrupolar or fully
symmetric irreps.  These are precisely the ones displaying Raman
activity, so we can restrict our attention just to Raman-active
modes when computing $\mu\plq$.

In this manuscript we consider three types of cubic crystals.
First, for C and Si in the diamond structure, the dynamical
charges vanish so that $e\pld$ and $\mu\pld$ vanish.  Of course
$e\pel$ also vanishes as the crystal is not piezoelectric.
However, $\mu\plq$ does not vanish, because the zone-center
optic mode has $T_2^+$ ($\Gamma_{25}^+$) symmetry and is
Raman-active.  The symmetry is such that
$\G_{1\a\b\c}=-\G_{2\a\b\c}=\gamma\,\varepsilon_{\a\b\c}$,
$Q\ptwo_{1\a\b\c}=-Q\ptwo_{2\a\b\c}=q\,\varepsilon_{\a\b\c}$
(where $\varepsilon$ is the fully antisymmetric tensor and ``1'' and
``2'' label the two atoms in the primitive cell). 
It is a short exercise to show that
$\mlone=\mltwo=0$ and $\mtr=-3\gamma q/2\Vc$.  In these
materials, therefore, $\mlone$ and $\mltwo$ have only
electronic contributions, while $\mtr$ has both electronic
and lattice-quadrupole contributions.

Second, we consider binary materials in the rocksalt or
cesium chloride structure.  The zone-center modes consist of
two copies of the $T_1^-$ ($\Gamma_{15}^-$) irrep, which is
not Raman-active.  In this case, $\mlone$, $\mltwo$ and
$\mtr$ all have contributions from electronic and
lattice-dipole terms only.

Third, we consider perovskite ABO$_3$ compounds.  Here the
zone-center modes comprise four copies of the IR-active $T_1^-$
($\Gamma_{15}^-$) irrep, plus one $T_2^-$ ($\Gamma_{25}^-$) irrep
that is neither IR nor Raman active. The situation is therefore
similar to the binary-compound case, with $\mlone$, $\mltwo$ and
$\mtr$ having electronic and lattice-dipole contributions only.
Two of the $T_1^-$ irreps correspond simply to displacements of
the A or B atom, while the other two correspond to particular
linear combinations of oxygen displacements.
The symmetry-mode treatment of the oxygen displacements in the perovskite
structure is detailed in Appendix \ref{app:oxygen}.

Note that more complex cubic crystals, such as spinels
and pyrochlores, may have both IR-active and Raman-active
zone-center modes.  In particular, any cubic crystal having
one or more free Wyckoff coordinates has Raman-active $A_1^+$
($\Gamma_1^+$) modes.  For such materials, $\mlone$, $\mltwo$
and $\mtr$ may all have contributions from all three terms in
the flexoelectric response.

\subsubsection{Pseudo-inverse of force-constant matrix and
force-pattern dependence}
\mylabel{sec:force-pattern}

Recall that the force-response tensors $\G$ and $T$ of
\equs{Ldef}{Tdef} need to be converted into displacement-response
tensors $\L$ and $N$ of \equs{Gdef}{Ndef} via the application of
a pseudo-inverse as in \equs{GKL}{NKT} respectively.  For the
piezoelectric response this is straightforward, because $\L$
obeys the acoustic sum rule, i.e., $\sum_I \L\bitbc=0$.
This reflects the fact that a uniform strain induces no net force
on an entire unit cell.  Unfortunately this is not true in general
for $T$, which describes the force response to a uniform
strain \textit{gradient}.  In general such a strain gradient is accompanied
by a stress gradient, i.e., a force density proportional to
$\nabla\cdot\boldsymbol{\sigma}$.  This means that
$\sum_I T\bitbcd \ne 0$ in general, and the definition and
application of the pseudo-inverse in \equ{NKT} is more subtle.

For the present purposes we can regard the force
$T\bitbcd$ induced by a strain gradient $\nu\bbcd$
as an ``external'' force $f\pext\bit$, and we have in general
that $b_\t\equiv\sum_I f\pext\bit \ne 0$.  We would like to
find a set of displacements $u\bit$ obeying
\beq
f\pext\bit - \sum\bjtp K_{I\t,J\t'} \, u\bjtp = 0 \;.
\eeq
We know this is not possible, however,
since $K$ obeys the ASR $\sum_I K_{I\t,J\t'}=0$,
so applying $\sum_I$ on the left-hand side yields $b_\t$.
Therefore, the best we can hope to do is to find a solution to
\beq
f\pext\bit - \sum\bjtp K_{I\t,J\t'} \, u\bjtp = b_\t\,w_I
\mylabel{eq:fbalance}
\eeq
instead, where the $w_I$ are a set of weights obeying
$\sum_I w_I=1$.  These weights describe the residual force pattern
that is left after the displacements $u\bit$ are applied, and we
have freedom to choose these as we wish.  For example, setting
$w_I=0$ except for $w_1=1$ would establish that
we seek a displacement pattern that makes the force vanish on all atoms
in the cell except atom 1.  A more natural choice is the ``evenly
weighted'' one given by $w_I=1/N$ for all $I$, which asks for
displacements that leave an equal residual force on every atom.
A third possibility is the mass-weighted choice $w_I=M_I /M_{\rm tot}$
where $M_{\rm tot}=\sum_I M_i$ is the total mass per cell.
This choice does affect the computation of the individual
FEC tensor components, because the dynamical charges $Q\pone$
appearing in \equ{muldT} depend on atom $I$, thus yielding a
different response to different displacements.
In Sec.~\ref{sec:results} we normally present our results for the
second and third of the choices discussed above (evenly weighted
or mass-weighted), which appear to be the most natural ones.

In Appendix~\ref{app:inverse} we explain how to define and compute
a pseudo-inverse $J^{[w]}_{I\t,J\t'}$ to $K_{I\t,J\t'}$ having the
desired property that
\beq
u\bit = \sum\bjtp J^{[w]}_{I\t,J\t'} \, f\pext\bjtp
\eeq
solves \equ{fbalance}.  The superscript $[w]$ appears on the
pseudo-inverse to emphasize that it is not unique, but depends
on the choice of force pattern embodied in the weights $w_I$.
We then use this pseudo-inverse $J^{[w]}$ in place of
$(K^{-1})$ in \equ{NKT} or (\ref{eq:muldT}).

The freedom in the choice of the force-pattern weights $w_I$
seems disconcerting at first sight, but we emphasize that any
\textit{physical} prediction of our theory is either independent
of this choice, or else determines it in an obvious way.  For
example, consider a \textit{static} deformation $\u(\rr)$ such as
that occurring in the beam-bending configuration discussed in
Sec.~\ref{sec:apxm}.  In this case, a combination of different
strain-gradient components $\nu\bbcd$ is present, and while the
net force per unit cell arising from just one of these components
may be non-zero, it must be canceled by those associated with
the other components.  Thus, individual FECs such as $g_{1111}$
and $g_{1122}$ in \equ{eff-FEC} may be force-pattern dependent,
but the effective coupling $g^{\rm eff}$ will not be.

Alternatively, consider the case of a crystal that is in static
equilibrium under the force of gravity, as for a sample sitting
on a tabletop.  A uniform strain gradient $\nu_{333}$ is
present because the force of gravity provides a downward external
force $f\pext$ along the vertical direction, compressing the
bottom of the sample more than the top.  The polarization
induced by this strain gradient is admittedly small, but so
is the strain gradient itself, and the ratio between these defines
a FEC.  Clearly this FEC should be computed using the mass-weighted
choice of force pattern, since gravity applies forces in proportion
to masses.  The mass-weighted choice is also appropriate to the
study of the dynamics of long-wavelength acoustic phonons, since
the force needed to accelerate an atom during its acoustic oscillation
is again proportional to its mass.

\subsection{FEC under different electric boundary conditions}
\mylabel{sec:fixd}

Up to now we have not been careful to distinguish quantities
defined at fixed electric field $\E$ from those defined
at fixed electric displacement field $D$. Most of our
calculations are performed under fixed-$D$ boundary conditions,
but experimental results are typically reported in terms of
fixed-$\E$ coefficients.  In this section we give the relationships
between the two kinds of quantities, which will be denoted with
superscripts ``$D$'' and ``$\E$'' to specify the type of electric
boundary conditions under which they are defined.

The relationship between the fixed-$\E$ and fixed-$D$ FECs follows
from
\bea
\md\babcd &=& \frac{dP\ba} {d\n\bbcd} \, \Bigg\vert_{D=0} \, \nn
          &=& \frac{\pa P\ba} {\pa \n\bbcd} \, \Bigg\vert_{\E=0} \, +
\, \frac{\pa P\ba} {\pa \E_{\l}} \,\frac{\pa \E_{\l}} {\pa \n\bbcd}
\,\Bigg\vert_{D=0}  \nn
&=&  \me\babcd - 4 \pi \, \chi_{\a\l} \, \frac{\pa P_{\l}} {\pa
\n\bbcd}
\,\Bigg\vert_{D=0}  \nn
&=& \me\babcd -4 \pi \, \chi_{\a\l} \, \md_{\l \b\c\d} \; .
\mylabel{eq:md-e}
\eea
In the third line above we introduce the (full lattice plus
electronic) dielectric susceptibility $\chi_{\a\l}=\pa
P\ba / \pa \E_{\l}$ and use $D=\E+4 \pi P$. Moving the
$4\pi\chi \md$ term to the left-hand side, this becomes
\beq
\epsilon^0_{\a\l} \, \md_{\l\b\c\d} = \me\babcd
\mylabel{eq:flexo-d2e}
\eeq
where $\epsilon^0_{\a\l} = \delta_{\a\l} + 4 \pi \chi_{\a\l}$ is
the static dielectric constant.

In the above derivation we assumed that the atoms could relax
in response to the applied strain gradient, arriving at \equ{flexo-d2e}.
The entire argument can be repeated for the frozen-ion FECs,
in which case $\chi$ and $\epsilon^0$ are replaced by
$\chi\pel$ and $\epsilon^\infty$ respectively, leading to
\beq
\epsilon^{\infty}_{\a\l} \, \mbard_{\l\b\c\d} = \mbare\babcd \; .
\mylabel{eq:flexo-frozen-d2e}
\eeq

Introducing the Born effective charge tensor
$\Ze_{I\a\tau}=\Vc (\pa P_\a/\pa u_{I\tau})|_{\E}$ and the
Callen effective charge tensor
$\Zd_{I\a\tau}=\Vc (\pa P_\a/\pa u_{I\tau})|_D
=(\epsilon_\infty^{-1})_{\a\b}\,\Ze_{I\b\tau}$, defined at
fixed $\E$ and $D$ respectively, we can relate
the zone-center force-constant matrices via
\beq
\Kd_{I\a,J\b} = \Ke_{I\a,J\b} + \frac{4 \pi} {\Vc} \,
\Ze_{I\a\l} \, \Zd_{J\b\l} \,.
\mylabel{eq:T2d-e}
\eeq
Similarly, the force-response internal strain-gradient tensors are related by
\beq
\Td_{I\a\b\c\d} = \Te_{I\a\b\c\d} -4 \pi \, \Ze_{I\a\l}\, \mbard_{\l
\b\c\d} \; .
\mylabel{eq:K2d-e}
\eeq

In the case of an isotropic dielectric tensor
$\epsilon^{\infty}\bab=\epsilon_\infty\,\delta\bab$, as occurs in cubic crystals,
we can make contact with \equ{Q1} by identifying
$Q\pone\biat=\Zd_{I\a\tau}=\epsilon_\infty^{-1}\,\Ze_{I\a\tau}$.
In the most general case, however,
it is unclear whether $Q\pone\biat$ and the higher moments
defined in \equr{Q1}{Q3} will transform as true tensors, and
the connection to the Callen charge would have to be reconsidered.


\subsection{Flexocoupling tensor}
\mylabel{sec:fcc}

Like for the FECs, there are many different notations
for the FCCs (flexocoupling coefficients)
in the literature, e.g., $2(\gamma + \eta)$ in
Refs.~[\onlinecite{catalan-jpcm04,catalan-prb05,eliseev-prb09,zhou-pb12}],
$h$ in Ref.~[\onlinecite{maranganti-prb09}], and even $f$ in
Ref.~[\onlinecite{yurkov-jetpl11}] related to unsymmetrized
strain. Here we follow
Refs.~[\onlinecite{zubko-armr13,yudin-unpub,ponomareva-prb12}]
in defining the FCC $f_{\a\d\b\c}$ as the coefficient in the
flexoelectric contribution
\beq
-\,\half f_{\a\d\b\c} \left( P\ba \, \frac{\pa \epsilon_{\b\c}}{\pa r\bd}
- \epsilon_{\b\c} \,\frac{\pa P\ba}{\pa r\bd} \right)
\label{eq:FCC}
\eeq
to the thermodynamic potential density.
Minimizing this energy functional leads to
\beq
g\pE_{\a\d\b\c} =  \chi_{\a\l}\, f_{\l\d\b\c} \; .
\eeq
Eqs.~(\ref{eq:sym-FEC}) and (\ref{eq:flexo-d2e}) imply that
\beq
g\pE_{\a\d\b\c} = \epsilon^0_{\a\l} \,
g\pD_{\l\d\b\c} \; .
\mylabel{eq:ge-d}
\eeq
We can also relate $f_{\a\d\b\c}$
to the FEC under fixed-$D$ boundary condition via
\beq
g\pD_{\a\d\b\c} =\frac{1}{4\pi}( \delta_{\a\l} -
\epsilon^{0 \; -1}_{\a\l}) \; f_{\l\d\b\c} \; .
\mylabel{eq:FECd-f}
\eeq
For high-$K$ materials it is reasonable to make the
approximation that $f_{\a\d\b\c} \simeq 4\pi\,g\pD_{\a\d\b\c}$.
(Note that while Gaussian units have been used in the derivations
here, the results for FECs and FCCs in Sec.~\ref{sec:results} are
converted to SI units for easier comparison with experimental
and previous theoretical results.)
In some previous works the FCCs have been obtained
either by deriving them from incomplete experimental
results\cite{zubko-armr13,yudin-unpub} or indirectly from
first-principles calculations on small supercells.\cite{ponomareva-prb12}
Once the FCCs have been obtained, they can
be used to obtain reasonable estimates for the room-temperature
FECs $g\pE_{\a\d\b\c}$ via Eqs.~(\ref{eq:ge-d}) and (\ref{eq:FECd-f}).

Finally, we note in passing that the (macroscopic
electrostatic) absolute deformation potentials $D^{\rm (macro)}$
of Refs.~\onlinecite{resta-prb90,resta_prb91erratum} can be related
to the FEC $g$ and the FCC $f$ via
\beq
D^{\rm (macro)}_{\a\d\b\c} = 4 \pi\, e\,  g_{\a\d\b\c}
\eeq
and
\beq
D^{\rm (macro)}_{\a\d\b\c} =  e \, (\delta_{\a\l}- \epsilon^{0 \; -1}_{\a\l}) \, f_{\l\d\b\c}
\mylabel{eq:fcc-adp}
\eeq
where $e$ is the electron charge.

\section{First-principles calculations}
\mylabel{sec:calculation}

Although we have laid out the formalism for the theory of
flexoelectricity above, expressing the FEC tensor in terms of more
elementary objects, it is still a challenge to
calculate this tensor from first principles.

In our previous work\cite{hong-prb11} we described how to calculate
the longitudinal frozen-ion component $\mbard\blone=\mbard\bl$ under 
fixed-$D$ boundary conditions\cite{hong-prb11-2} from first-principles.  
Here we first review supercell cell calculations and
discuss how to extend them to obtain the $T$ tensor
elements needed for the lattice flexoelectric response. We also
show how to carry out similar supercell calculations, but in a
rotated frame, to obtain the corresponding $\m\bltwo$ components.
We then provide the details of the first-principles calculations, and
briefly present some computed information about the atomic
cores and about the ground-state properties of the crystals that
will be needed later.

\subsection{Supercell calculations in original Cartesian frame}
\mylabel{sec:orig-frame}

\begin{figure}
  \begin{center}
    \includegraphics[width=3.3in]{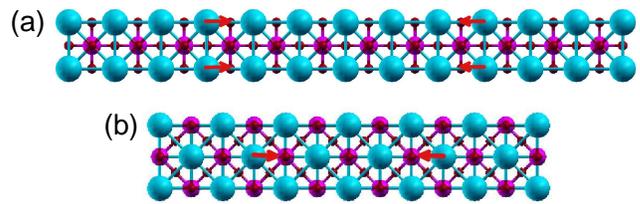}
  \end{center}
  \caption{\mylabel{fig:model}
  (Color online) Original (a) and 45$^\circ$-rotated
  (b) supercells used for calculations on ABO$_3$ perovskites.
  Arrows (red) denote the displacement of A atoms consistent
  with fixed-$D$ boundary conditions. Large ball is
  A atom, medium is B atom, and small is O atom.
    }
\end{figure}

Fig.~\ref{fig:model}(a) illustrates the supercell
that we introduced in Ref.~\onlinecite{hong-prb11} in order
to compute $\mu\pel_{1111}$.  For each type of atom, we move two
planes of these atoms, located approximately 1/4 and 3/4 along the
supercell long dimension, by equal and opposite amounts, as
illustrated in the figure.  We do this
in order that the electric field between these displaced
planes should vanish; since the polarization also vanishes there,
this corresponds to fixed-$D$ boundary conditions.  As a result,
we obtain a very rapid spatial convergence (locality) of the induced
charge distribution $f_{l\t}(\rr)$ of \equ{f}, and of the induced
forces reflected in the force-constant elements $\Phi^{lIJ}_{\t\t'}$
of \equ{Phi}, as illustrated in Fig.~\ref{fig:check-fixd}.
(For details of these calculations, see Sec.~\ref{sec:details}.)
Displacing only a single plane of atoms with $\E=0$ boundary
conditions on the entire supercell would set up a macroscopic
local $\E$-field even far from the displaced plane leading to
oscillations in $f_{l\t}(\rr)$ and the corresponding forces,
making it difficult or impossible to calculate the needed
spatial moments of $f_{l\t}(\rr)$ and of the induced forces.

\begin{figure}[t]
\begin{center}
\includegraphics[height=1.8in]{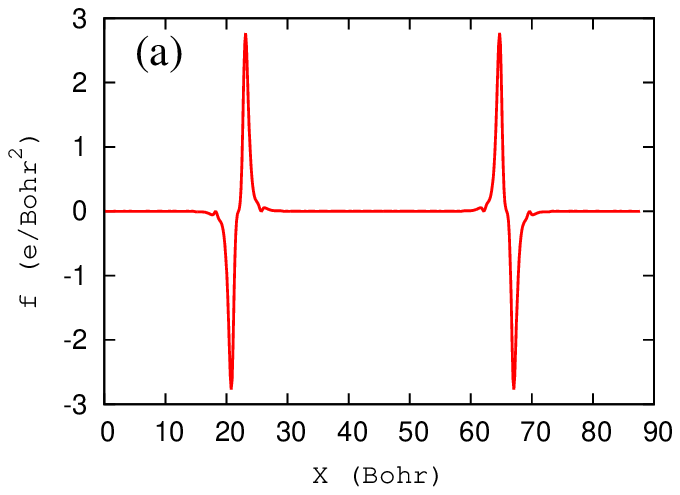}
\includegraphics[height=1.8in]{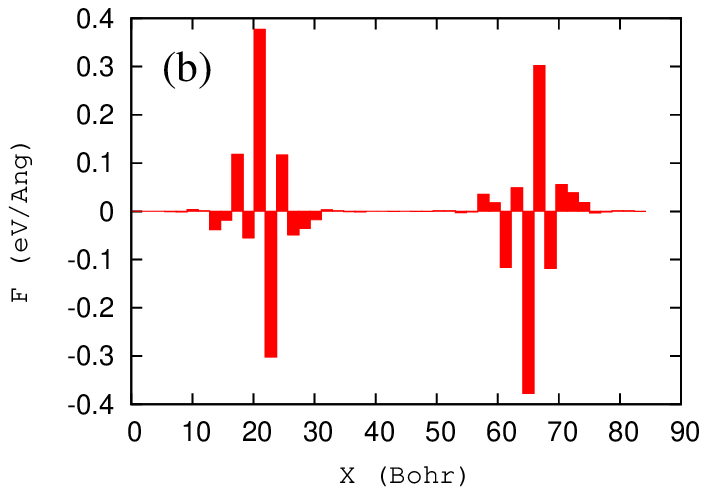}
\end{center}
\caption{\mylabel{fig:check-fixd}
(Color online) Change of charge density distribution (a) and force
distribution (b) in \STO\ supercell (original frame) at fixed
$D$.}
\end{figure}

 From finite differences of the computed charge densities with
small positive and negative displacements, repeated for each type of
atom $I$, we calculate $Q\pone_{I11}$ and
$Q\pthr_{I1111}$ via \equs{Q1}{Q3} respectively.  We emphasize
again that these are fixed-$D$ quantities by definition, and so are
given correctly by the configuration of Fig.~\ref{fig:model}.

At the same time, we compute the forces on all the atoms in
the supercell as illustrated in Fig.~\ref{fig:check-fixd}(b),
and use these to construct the force-constant elements needed
for computing $T\pD_{I1111}$ from \equ{T1}.  In practice this
works as follows.
Let $i$ denote the atom in the supercell for which we
want to compute $T_{1111}$, and let $j$ run over other atoms in the
supercell.  Imagine that there is a uniform strain gradient
causing displacements
\beq
u_{jx}=\frac{1}{2}\,\nu_{xxx}\,(\Delta x_{ij})^2
\mylabel{eq:ujx}
\eeq
in the vicinity of atom $i$, where $\Delta x_{ij}=x_j-x_i$.
The total force on atom $i$ would then be
\beq
f_{ix}=\sum_j F_{ix}^{(jx)}\,u_{jx}
\mylabel{eq:ijx}
\eeq
where $F_{i\a}^{(j\b)}$ is the force induced on atom $i$ in
direction $\a$ by a displacement of atom $j$ in direction $\b$.
Using the definition that $T_{i,xxxx}=f_{ix}/\nu_{xxx}$ and
substituting \equ{ujx} into (\ref{eq:ijx}), we get
\beq
T\pD_{i,xxxx}= \frac{1}{2} \sum_j F_{ix}^{(jx)}\,(\Delta x_{ij})^2 \;.
\eeq
Note, however, that $F_{i\a}^{(j\b)}$ is just minus the
zone-center force-constant matrix of the supercell, which is
symmetric under interchange of indices, so the above can be
rewritten as
\beq
T\pD_{i,xxxx}= \frac{1}{2} \sum_j F_{jx}^{(ix)}\,(\Delta x_{ij})^2
\;.
\mylabel{eq:Tsup}
\eeq

\equ{Tsup} is the formula that we use to calculate $T\pD_{I1111}$ in
practice.  That is, rather than displace other atoms and compute
the force on atom $i$, we displace atom $i$ and compute the forces
on other atoms, then calculate the the second moment of these
forces from \equ{Tsup}.  The sum is truncated when the distance
$|\Delta x_{ij}|$ approaches half the distance to the next
plane of displaced atoms (i.e., $\sim$1/4 of the supercell long
dimension).  For large enough supercells, this is already in
the region in which the $F_{jx}^{(ix)}$ have essentially vanished
(i.e., see Fig.~\ref{fig:check-fixd}(b)),
so that the sum is well converged.  Note that \equ{Tsup} is
essentially the same as \equ{T1}, but adapted to practical
supercell calculations.

We also carry out calculations in which the plane of
atoms is displaced in the transverse $y$ direction, i.e., vertically
in Fig.~\ref{fig:model}, and compute the $y$ forces on the other
atoms in the cell.  This is not useful for computing moments of
the $Q$ tensors, but it allows us to compute the $T\pE_{I2211}$
(later presented as $T\pE_{I1122}$)
which are eventually needed to compute $\m\pld\blt$, in a manner
entirely analogous to the $T\pD_{I1111}$ calculation.  Note,
however, that the calculation is carried out at fixed (vanishing)
$\E_y$ in this case, so the resulting quantity is to be interpreted
as a fixed-$\E$ one, as indicated by the superscript on
$T\pE_{I2211}$.

For the case of oxygen atoms in the perovskite structure,
$T\pD_{I1111}$ and $T\pE_{I2211}$ are computed as above for
$I$ = O1, O2 and O3, and then converted into the symmetry-mode
representation ($\xi$=3,4) as described in Appendix~\ref{app:oxygen}.

\subsection{Supercell calculations in rotated frame}
\mylabel{sec:rotate-frame}

The calculations described above are sufficient to compute the
$Q_{I1111}$ and $T\pD_{I1111}$
tensor components needed to compute the electronic and
lattice parts of $\mlone$ of \equ{mlone}, but not
$\mltwo$ of \equ{mltwo}.
In order to calculate the latter, we introduce the rotated frame
shown in the  Fig.~\ref{fig:model}(b)
and calculate the longitudinal FEC in this rotated frame.

We label the FEC in the original frame as $\m\babcd$ and in
rotated frame as $\mpri\babcd$. These are related by
applying the rotation matrix
\beq
R(\theta) =
\begin{pmatrix}
\cost & -\sint & 0 \cr
\sint &  \cost & 0 \cr
0 & 0 & 1
\end{pmatrix}
\eeq
with $\theta$=45$^{\circ}$ four times,
\beq
\mpri_{\a'\b'\c'\d'} = \sum_{\a\b\c\d} R_{\a'\a}\, R_{\b'\b}\,
     R_{\c'\c}\, R_{\d'\d} \, \m_{\a\b\c\d} \; ,
\mylabel{eq:frot}
\eeq
giving
\bea
\mpri\bl    &=& \half(\m\bl + \m\btr) + \m\btm \; ,
\mylabel{eq:m1111} \\
\mpri\btr   &=& \half(\m\bl + \m\btr) - \m\btm \; ,
\mylabel{eq:m1}\\
\mpri\btm   &=& \half(\m\bl - \m\btr) \;.
\mylabel{eq:m1221}
\eea
Referring to \equr{mlone}{mtr},
note that $\m'_{1111}-\mu_{1111}=(\mltwo-\mlone)/2$,
confirming that $\Delta=\mltwo-\mlone$ is a measure of anisotropy as
was discussed there.  From \equ{m1111} it follows that
\beq
\mltwo=2\m'_{1111} - \m_{1111}  \;.
\mylabel{eq:rotation-L}
\eeq
It is therefore straightforward to obtain the missing
FEC component $\mltwo$ once $\m'_{1111}$ has been
calculated.

To obtain $\mu\ppel_{1111}$, we compute
${Q'}\pthr_{I1111}=Q\pthr_{I,x'x'x'x'}$
for each atom $I$ in the rotated supercell just as we did for
$Q\pthr_{I1111}=Q\pthr_{I,xxxx}$ in the original cell.  However, as
explained in Appendix~\ref{app:oxygen}, for the oxygen atoms in
perovskites we have to compute $Q\pthr_{\O1,x'y'x'x'}$ as well.
Since $Q\pthr_{\O1,x'y'x'x'}=-Q\pthr_{\O2,x'y'x'x'}$, a convenient
way to do this is to move atoms O1 and O2 by equal and opposite
amounts along $y'$, thus preserving the $\E_y=D_y=0$ boundary conditions
as was done for other displacements.  For the lattice part,
we similarly need the $T\pD$ tensors in the rotated frame.
The $T\pD_{x'x'x'x'}$ matrix elements are computed similarly as
for the original supercell, except that for oxygens in perovskites
we also need $T\pD_{O1,y'x'x'x'}$ (see Appendix~\ref{app:oxygen}).
Again, this requires a displacement of O1 along $y'$ (or better,
equal and opposite displacements of O1 and O2 along $y'$), with
the $x'$-second-moments of the $x'$-forces on the other atoms
obtained in the same way as for $x'$ displacements.

\subsection{Details of the calculations}
\mylabel{sec:details}

The calculations have been performed within density-functional
theory.  We used the local-density approximation~\cite{perdew-prb81}
for C, Si, MgO, NaCl, CsCl and \STO, and the generalized gradient
approximation~\cite{wu-cohen-prb06} for \BZO, \BTO\ and \PTO.
We used SIESTA~\cite{soler-jpcm02} package for the calculations.
Norm-conserving pseudopotentials were used, with semicore shells
included for for Ti (3s3p3d), Ba (5s5p), Zr (4p4d), Pb (5d),
Sr (4s4p), and Cs (5s5p).
In all cases the $k$-space mesh was chosen to correspond to a
12\,\AA\ cut-off\cite{moreno-prb92} while the real-space integrals
were carried out on an $r$-space mesh corresponding to a
450\,Ry cutoff.\cite{soler-jpcm02}
Supercells were built from 12 unit cells for CsCl and
perovskites in the original frame and 6 cells in the $45^{\circ}$-rotated frame
(see Fig.~\ref{fig:model}). For C, Si, MgO and NaCl, we used 8 conventional
cells in in original frame and 4 cells in the $45^{\circ}$-rotated frame.
Atomic displacements of 0.04\,\AA\ were used in our calculations. In order to
reduce the anharmonic effect, two calculations were performed, one with negative
displacement and the other one with positive displacement.

For the cubic perovskite structure ABO$_3$, atoms A and B have
the cubic symmetry, but the individual O atom has tetragonal
symmetry, not the cubic symmetry. In our calculation, we chose
to use ``mode coordinate'' for perovskites in which two oxygen
modes have the cubic symmetry. Please refer to
Appendix~\ref{app:oxygen} for the details.

\subsection{Ground-state properties of materials}

In order to calculate the FEC, we need to obtain some basic
properties of our materials of interest, including the lattice
constant ($a$), optical dielectric
constant ($\epsilon^{\infty}$), static dielectric constant
($\epsilon^0$), and Born
effective charges.  These are summarized in
Table~\ref{tab:properties1}.

The optical dielectric constants are obtained as $\epsilon^{\infty} =
Z^{\E} / Q^{(1)}$, where $Q^{(1)}$ is the Callen charge as given in
Table~\ref{tab:rawdata}.  (For C and Si, $\epsilon^{\infty}=
\epsilon^0$ and we do not compute $\epsilon^{\infty}$ explicitly.)
The Born effective charges $Z^{\E}$ are obtained by finite differences,
e.g., by displacing each atom slightly and calculating the induced
Berry-phase polarization.\cite{kingsmith-prb93}

The theoretical static dielectric constant is obtained via
$\epsilon^0 = \mu^{\E}/\mu^D$ following \equ{flexo-d2e}. We also
checked that these results are exactly the same as obtained by
the usual method based on the effective-charge and force-constant
tensors (see, e.g., Ref.~\onlinecite{wu-prb05}).  However,
we do not attempt to compute $\epsilon^0$ for
\BTO, \PTO\ and \STO, because
our calculations are for the reference cubic paraelectric
structure, which is either unstable (\BTO\ and \PTO) or nearly
unstable (\STO) to the formation of a ferroelectric ground state.
For this reason, a direct calculation of $\epsilon^0$
would give negative or extremely large formal values, leading to
nonphysical results for $\mu^{\E}$ and other quantities in subsequent
calculations.  Instead, we have chosen to adopt room-temperature
experimental values for $\epsilon^0$ for these materials, as
given in Table~\ref{tab:properties1}, and we use these to
predict meaningful values of $\mu^{\E}$ at room temperature.
Our rationale for this approach was explained in more detail
at the end of Sec.~\ref{sec:intro}.

\begin{table}
\caption{\mylabel{tab:properties1}
Lattice constant $a$ (of conventional cell\cite{explan-cell}),
optical dielectric constant $\epsilon^{\infty}$, theoretical
(zero-temperature) and experimental (room-temperature) static
dielectric constant $\epsilon^0$, and Born effective charge
$Z^{\E}$ for
materials considered in this study.
}
\begin{ruledtabular}
\begin{tabular}{lrrccrrr}
& \multicolumn{1}{c}{$a$} &
\multicolumn{1}{c}{$\epsilon^{\infty}$} & 
\multicolumn{1}{c}{$\epsilon^0$} &
\multicolumn{1}{c}{$\epsilon^0$} & &
\multicolumn{1}{c}{$Z^{\E}$}  \\
& \multicolumn{1}{c}{(Bohr)}
&
& \multicolumn{1}{c}{theo.}
& \multicolumn{1}{c}{expt.\tablenotemark[1]}
& &
  \multicolumn{1}{c}{($e$)} \\
\hline
 C     & 6.69  &--   &--  & 5.5      & C     & 0\\
 Si    & 10.22 &--   &--  & 11.9     & Si    & 0\\
 MgO   & 7.82  &3.0  &7.8 &9.8     & Mg    & 1.89     \\
       &       &     &    &        & O     & --1.89     \\
 NaCl  & 10.66 &2.4  &6.1 &5.6     & Na    & 1.07       \\
       &       &     &    &        & Cl    & --1.07   \\
 CsCl  & 7.45  &3.2  &6.4 &7.2     & Cs    & 1.36       \\
       &       &     &    &        & Cl    & --1.36   \\
  BZO  & 7.92  &5.0  &57.1&15      & Ba    & 2.84     \\
       &       &     &    &        & Zr    & 6.26     \\
       &       &     &    &        & O$_1$ &--5.04    \\
       &       &     &    &        & O$_3$ &--2.03    \\
  BTO  & 7.52  &6.8  &--  &2300    & Ba    & 2.78     \\
       &       &     &    &        & Ti    & 7.50     \\
       &       &     &    &        & O$_1$ & --6.04   \\
       &       &     &    &        & O$_3$ & --2.12   \\
  PTO  & 7.43  &8.8  &--  &134     & Pb    & 3.93     \\
       &       &     &    &        & Ti    & 7.21     \\
       &       &     &    &        & O$_1$ & --6.03   \\
       &       &     &    &        & O$_3$ & --2.55   \\
   STO & 7.31  &6.3  &--  &310     & Sr    & 2.52     \\
       &       &     &    &        & Ti    & 7.47     \\
       &       &     &    &        & O$_1$ & --5.89   \\
       &       &     &    &        & O$_3$ & --2.05   \\
\end{tabular}
\tablenotetext[1]{Room-temperature experimental values:
  MgO and NaCl, Ref.~\onlinecite{mgo-e};
  CsCl, Ref.~\onlinecite{cscl-e};
  BZO, Ref.~\onlinecite{bzo-e};
  BTO, Ref.~\onlinecite{ma-apl06};
  PTO, Ref.~\onlinecite{pto-e};
  STO, Ref.~\onlinecite{sto-e}.}
\end{ruledtabular}
\end{table}

\subsection{Rigid-core corrections for pseudopotentials}

\begin{table}
\caption{
Rigid core corrections (RCCs), in e\,bohr$^2$. Pseudopotentials
O1 and O2 are used for MgO and perovskites respectively.
\mylabel{tab:rcc}
}
\begin{ruledtabular}
\begin{tabular}{lrlrlr}
& \multicolumn{1}{c}{RCC}
&& \multicolumn{1}{c}{RCC}
&& \multicolumn{1}{c}{RCC} \\
\hline
C  & $-$0.19 & Si & $-$2.93 & Cs & $-$14.58 \\
O1 & $-$0.06 & Cl & $-$1.67 & Ba & $-$13.34 \\
O2 & $-$0.09 & Ti & $-$0.84 & Pb & $-$15.46 \\
Na & $-$6.64 & Sr & $-$5.41 \\
Mg & $-$4.85 & Zr & $-$7.72
\end{tabular}
\end{ruledtabular}
\end{table}

Our previous work\cite{hong-prb11} showed that the frozen-ion
FEC $\m\pel$ is dependent on the choice of pseudopotential.
To see why, consider a
model in which each cation or anion is represented by a spherically
symmetric charge $\rho_i(r)$ that displaces rigidly as a unit.
A brief calculation shows that 
\beq
Q\pthr_i=\int d^3r\,x^3\,
(-\partial_x \rho_i(\rr))= 4\pi\int dr\,r^4\,\rho_i(r) \; .
\mylabel{eq:q3neg}
\eeq
It can be seen that $Q\pthr$ and hence $\mu\pel$ should depend on
the treatment of of the core density and the pseudopotential
construction.  The difference between $Q^{(3,{\rm AE})}$ (all-electron)
and $Q^{(3,{\rm PS})}$ (pseudopotential) can be corrected by
introducing a ``rigid core correction" (RCC)
\beq
Q^{(3,{\rm  RCC})}_i=
4\pi\int dr\,r^4\,[\rho^{\rm AE}_i(r)-\rho^{\rm PS}_i(r)]
\eeq
for each atom type $i$ by using the densities from free-atom AE and
PS calculations, and then adding these $Q^{(3,{\rm RCC})}$ corrections
to the $Q^{(3,{\rm PS})}$ values calculated in our supercells.

In order to obtain accurate FECs, we calculate this RCC for
all elements appearing in our selected materials, as shown in
Table~\ref{tab:rcc}. It can be seen that the RCC tends to be
large for large-radius atoms like Cs, Ba and Pb, even though their
semicore shells are included in the valence in the pseudopotential
construction.  From these values it is clearly essential
to include the RCC for elements with large radius if one wants
to compute the FxE response accurately.


\section{Results}
\mylabel{sec:results}

In this section, we first present the basic charge-moment
tensors $Q$ and force-response tensors $T$ as they are extracted
from our supercell calculations.  We then combine these to obtain
the longitudinal FECs at fixed $D$, as well as the longitudinal
flexocoupling coefficients, for all of the materials considered.
Up to this point, the results do not depend on current-response
terms $\m\plqJ$ and $\m\pelJ$ in \equs{mulq}{muel}, but to go
further we then make the uncontrolled assumption that these two
quantities vanish in order to get a rough idea of the behavior
of the full FEC tensors at fixed $D$ and $\E$.

\subsection{Calculated charge-moment and force-response tensors}
\mylabel{sec:raw-results}

As described in Secs.~\ref{sec:orig-frame} and
\ref{sec:rotate-frame}, we carry out calculations on supercells
extended along $x$ with small $x$ displacements to obtain
$Q\pone$, $Q\pthr\blone$, and $T^D\blone$ (all at fixed $D$).
We also calculate $T^{\E}_{1122}$ by applying small $y$
displacements (at fixed-$\E$ boundary conditions) instead.  Then,
we carry out similar calculations on 45$^\circ$-rotated supercells
to obtain $Q\pthr\bltwo$ and $T\bltwo$ as well.  The treatment
of the oxygen displacements in perovskites require special care
as described in Sec.~\ref{sec:rotate-frame} and Appendix~\ref{app:oxygen}.

The results are presented in Table~\ref{tab:rawdata}.  The
RCCs from Table~\ref{tab:rcc} have been included in the calculation
of the $Q\pthr$ moments.  From Table~\ref{tab:rawdata}, it can
be seen that a modified acoustic sum rule has to be used for the
mode variables in ABO$_3$ perovskites, namely,
$Q\pone_{\rm A} +  Q\pone_{\rm B}  + Q\pone_{\xi_3} +
\sqrt{2} Q\pone_{\xi_4} = 0$.
Almost all $Q\pthr$ values are seen to be negative, which can be understood
heuristically from the rigid-ion model of Eq.~(\ref{eq:q3neg}),
in which the positive nuclear charge at $r=0$ makes no contribution
so that the electronic charge gives an unbalanced negative
contribution to $Q\pthr$.  The only exceptions are for the
$\xi_4$ modes in the perovskites, where the oxygen
motion along $y'$ is involved so that the sign is less
intuitive.  In the ABO$_3$ perovskites, the
A atom makes a significant contribution to all $Q$ and $T$
tensors. Also note that oxygen mode $\xi_4$ contributes only very
weakly to $T^D\blone$, while $\xi_3$ contributes strongly.

\begin{table}
  \caption{ \mylabel{tab:rawdata}
  $Q^D$ and $T$ tensors computed from first-principles
  calculations. Units: e for $Q\pone$;
  e\,Bohr$^2$ for $Q\pthr$; eV for $T$.
  The RCCs from Table~\ref{tab:rcc} are included in the $Q\pthr$
  moments.}
\begin{ruledtabular}
\begin{tabular}{lrrrrrrr}
& &
\multicolumn{1}{c}{$Q\pone$} &
\multicolumn{1}{c}{$Q\pthr\blone$} & 
\multicolumn{1}{c}{$Q\pthr\bltwo$} &
\multicolumn{1}{c}{$T^D\blone$} & 
\multicolumn{1}{c}{$T^D\bltwo$}  &
\multicolumn{1}{c}{$T^{\E}_{1122}$} \\  

\hline

 C     & C     &  0     & --13.2 & --25.4  &  0    &  0     & 0         \\
 Si    & Si    &  0     & --30.9 & --54.9  &  0    &  0     & 0        \\
 MgO   & Mg    &  0.63  & --13.7 & --23.3  &  37.5 &  65.4  & 16.3       \\
       & O     & --0.63 & --12.8 & --15.8  &  41.5 &  30.4  & 20.1         \\
 NaCl  & Na    &  0.45  & --10.1 & --24.2  &  10.5 &  11.8  & 30.8         \\
       & Cl    & --0.45 & --27.6 & --30.6  &  14.0 &  5.4   & -3.0       \\
 CsCl  & Cs    &  0.43  & --64.1 & --72.3  &  11.6 &  10.3  & 16.3        \\
       & Cl    & --0.43 & --30.2 & --40.1  &  8.1  &  11.6  & 20.1      \\
  BZO  & Ba    &  0.57  & --75.5 & --63.3  &  18.8 &  42.2  & 11.1  \\
       & Zr    &  1.25  & --30.5 & --63.5  &  63.9 &  37.6  &  5.3  \\
       &$\xi_3$& --1.01 & --34.7 & --132.0 &  73.4 &  35.6  &  6.2  \\
       &$\xi_4$& --0.57 & --11.7 &   67.9  &   1.8 &  10.3  &  7.5  \\
  BTO  & Ba    &  0.40  & --79.2 & --65.7  &  24.1 &  52.9  & 13.1 \\
       & Ti    &  1.11  & --14.9 & --46.0  &  46.0 &  40.2  &  7.6 \\
       &$\xi_3$& --0.89 & --29.2 & --109.4 &  45.2 &  36.4  & 14.4 \\
       &$\xi_4$& --0.44 & --8.4  &   52.4  &   9.2 &  17.9  & 15.6 \\
  PTO  & Pb    &  0.44  & --73.7 & --66.7  &  16.1 &  38.0  &  6.1  \\
       & Ti    &  0.83  & --23.8 & --49.5  &  44.0 &  38.6  &  3.5  \\
       &$\xi_3$& --0.69 & --24.5 & --135.8 &  52.0 &  25.6  & 13.4  \\
       &$\xi_4$& --0.41 & --12.9 &   69.6  &   4.9 &  20.3  & 12.0  \\
   STO & Sr    &  0.39  & --57.5 & --43.5  &  17.0 &  35.7  &  8.4  \\
       & Ti    &  1.20  & --16.2 & --45.0  &  52.3 &  38.9  &  3.0  \\
       &$\xi_3$& --0.92 & --28.6 & --44.0  &  68.7 &  13.1  & 15.7  \\
       &$\xi_4$& --0.47 &  --9.5 &   7.7   &   3.6 &  18.2  & 12.0  \\
\end{tabular}
\end{ruledtabular}
\end{table}

\subsection{Longitudinal contribution to flexoelectric response}
\mylabel{sec:result-l}

\begin{table*}
\caption{
Longitudinal components of FEC tensor at fixed $D$.
$\mlone=\m\bl$, $\mltwo=\m\btr + 2 \m\btm$, and anisotropy
$\Delta=\mltwo -\mlone$. Units: pC/m.
\mylabel{tab:fixd1}
}
\begin{ruledtabular}
\begin{tabular}{lrrrrrrrrrrrr}
& & & 
\multicolumn{5}{c}{Even force}  &
\multicolumn{5}{c}{Mass-weighted force}  \\
\cline {4-8} \cline{9-13} 
& 
\multicolumn{1}{c}{$\mu\pel\blone$}  &
\multicolumn{1}{c}{$\mu\pel\bltwo$}  &
\multicolumn{1}{c}{$\mu\pld\blone$}  &
\multicolumn{1}{c}{$\mu\pld\bltwo$}  &
\multicolumn{1}{c}{$\mlone$}  &
\multicolumn{1}{c}{$\mltwo$}  &
\multicolumn{1}{c}{$\Delta$}  &
\multicolumn{1}{c}{$\mu\pld\blone$}  &
\multicolumn{1}{c}{$\mu\pld\bltwo$}  &
\multicolumn{1}{c}{$\mlone$}  &
\multicolumn{1}{c}{$\mltwo$}  &
\multicolumn{1}{c}{$\Delta$}  \\
\hline
C    & --175.4  & --163.5 &     0&    0& --175.4 &--163.5  &11.9  &     0 &      0&--175.4&--163.5 &11.9  \\
Si   & --106.0  & --90.8  &     0&    0& --106.0 &--90.8   &15.2  &     0 &      0&--106.0&--90.8  &15.2  \\
MgO  &    --111.7& --164.8& --5.7& 50.5&  --117.4& --114.4 &   3.0& --29.2&   22.0& --140.8&--142.8&--2.0   \\
NaCl &     --62.8&  --91.2& --9.0& 16.3&   --71.8& --74.9  &--3.1 &    4.2&   25.6&  --58.6&--65.6 &--7.0  \\
CsCl &    --115.2& --137.4&   5.8&--2.2&  --109.4&--139.6  &--30.2 &--13.2 &--23.3 & --128.4&--160.6  & --32.2
\\
BZO  &    --154.8& --194.1&  29.3& 34.7&  --125.5& --159.3 &--33.8& --35.2& --18.1& --190.0&--212.2& --22.2  \\
BTO  &    --155.8& --199.7&  10.5& 9.8 &  --145.3& --189.9 &--44.6&  --5.4&  --9.4& --161.2&--209.1&--47.9 \\
PTO  &    --165.7& --224.1&  16.5& 23.1&  --149.2&  --201.0&--51.8& --31.4& --29.6& --197.1&--253.7&--56.6 \\
STO  &    --143.9& --160.9&  24.8& 22.3&  --119.1& --138.6 &--19.5& --12.1& --11.6& --156.1&--172.4&--16.3 \\
\end{tabular}
\end{ruledtabular}
\end{table*}

The longitudinal FEC constants $\mlone^D$ and $\mltwo^D$, and
their electronic and lattice contributions, are presented in
Table~\ref{tab:fixd1}, following the definitions in
\equs{mlone}{mltwo}.  The electronic (frozen-ion) response is
obtained from \equ{muel} (recall that $\m\pelJ$ does not contribute
to the longitudinal response), and is seen to be negative in all
cases, as expected from the sign of the $Q\pthr$ contributions.
The lattice quadrupole contribution $\m\plq$ of \equ{mulq} vanishes
for all of our compounds except for C and Si, where it only
makes a contribution to the transverse component $\mtr$, as discussed
in the next subsection.
Regarding the lattice dipole contributions $\m\pld$, which
vanish for C and Si, these are computed from \equ{muldT}
using the information in Table~\ref{tab:rawdata} together with the
computed zone-center force-constant matrices and their
pseudo-inverses.

As emphasized in Sec.~\ref{sec:force-pattern},
the lattice contributions to the FECs
depend on the force pattern applied on each atom in the unit
cell. Here we chose two different force patterns, either evenly
weighted on all atoms, or else weighted according to the atomic
mass.  From Table~\ref{tab:fixd1} we can see that the total fixed-$D$
FECs are comparable for all of our selected materials.  The lattice
contribution, and therefore the total FEC, is force-pattern
dependent. However,  the anisotropy ($\Delta=\mltwo - \mlone$)
\cite{yudin-prb12} shows hardly any force-pattern
dependence, and the
anisotropy is much larger in the perovskites than in the
elemental and binary materials. It can be seen that C, Si, MgO,
NaCl have a nearly isotropic behavior. From this table, we can also see that 
at fixed $D$ the lattice contribution is much smaller than the
electronic contribution.
For the even force pattern, the lattice contributions are positive,
which reduces the total FEC. However, under the mass-weighted force
pattern, the lattice contributions are negative, enhancing the 
total FEC. Therefore, the FECs under the mass-weighted force pattern
are larger (more negative) than those under the evenly-weighted force
pattern.

We next present the results for the flexocoupling coefficients
(FCCs) defined in Sec.~\ref{sec:fcc}, which can be regarded as
fundamental materials properties because they are not strongly
temperature-dependent.\cite{ponomareva-prb12}  The FCCs can easily
be obtained from our fixed-$D$ FECs via Eq.~(\ref{eq:FECd-f}).
To evaluate this equation we use the theoretical dielectric
constants $\epsilon^0$ in Table~\ref{tab:properties1} for MgO,
NaCl, CsCl and BZO, while using the experimental $\epsilon^0$
values for other materials.

\begin{table}
\caption{Longitudinal flexocoupling coefficients. $f_{\rm L1}=f_{1111}$,
$f_{\rm L2}= f_{1122} + 2 f_{1221}$, and anisotropy $\Delta=f\bltwo
-f\blone$. Units: V.
\mylabel{tab:fcc}
}
\begin{ruledtabular}
\begin{tabular}{lrrrrrc}
&
\multicolumn{3}{c}{Even force}  &
\multicolumn{3}{c}{Mass-weighted force}  \\
\cline {2-4} \cline{5-7}
&
\multicolumn{1}{c}{$f_{\rm L1}$}  &
\multicolumn{1}{c}{$f_{\rm L2}$}  &
\multicolumn{1}{c}{$\Delta$}  &
\multicolumn{1}{c}{$f_{\rm L1}$}  &
\multicolumn{1}{c}{$f_{\rm L2}$}  &
\multicolumn{1}{c}{$\Delta$}  \\
\hline
C    & --19.8  &--18.5  &  1.3    &  --19.8  &--18.5  &1.3 \\
Si   & --12.0  &--10.3  &  1.7    &  --12.0  &--10.3  &1.7 \\
MgO  & --15.2  & --14.8 &   0.4   & --18.2   & --18.5 &--0.3 \\
NaCl & --9.7   & --10.1 & --0.4   & --7.9    & --8.9  &--1.0 \\
CsCl & --14.6  & --18.7 & --4.1   & --17.2   & --21.5 &--4.3\\
BZO  & --14.4  & --18.3 & --3.9   & --21.8   & --24.4 &--2.6 \\
BTO  & --16.4  & --21.5 & --5.1   & --18.2   & --23.6 &--5.4 \\
PTO  & --17.0  & --22.9 & --5.9   & --22.4   & --28.9 &--6.5 \\
STO  & --13.5  & --15.7 & --2.2   & --17.7   & --19.5 &--1.8 \\
\end{tabular}
\end{ruledtabular}
\end{table}

The FCCs for our various materials under the two different force patterns
are presented in Table~\ref{tab:fcc}. It can be seen that
$|f_{\rm L1}| \equiv |f_{1111}|$ is in the range of $[10V, 20V]$, which is slightly
larger than the value $[1 V, 10 V]$ in a previous theoretical
estimate.\cite{yudin-unpub}
The FCC is slightly larger under the mass-weighted force pattern than
under the evenly-weighted one, and $f\bltwo$ is also larger
than $f\blone$. Interestingly,
the FCCs are all of roughly comparable size for all of the
materials, reinforcing the picture that they constitute
fundamental materials properties that are hardly affected by
the large static dielectric constants present in some materials.
The anisotropy of the FCCs is also nearly
force-pattern independent, just like
for the FECs. From Eq.~(\ref{eq:fcc-adp}) we calculate
$D^{\rm (macro)}_{\rm L1} = 11.0$ eV, close to the result (12.0 eV) in
Ref.~(\onlinecite{resta-prb90}).

Zubko \textit{et al.}\cite{zubko-armr13} summarized the available
FEC data in their recent review paper and converted them to FCCs
by using $f \simeq 4\pi g / \epsilon^0$.
They found that the FCCs change substantially
(including in sign) in different materials.  The discrepancy relative
to our results may result in part because their analysis rests on
experimental data for the effective beam-bending FECs
and they also include surface effects.

Ponomareva \textit{et al.}\cite{ponomareva-prb12} calculated the
FCCs for \BSTO\ from first principles by introducing periodic
strain gradients in small
supercells. Their results show that the FCC is indeed a fundamental
quantity that is only weakly dependent on the temperature and thickness
for \BSTO\ films. However, $f_{1111}$ is about 5\,V for \BSTO,
smaller than our value presented in Table~\ref{tab:fcc}, and with
opposite sign. One possible reason could be that their supercells
were too small; other previous work has shown that
the FEC is very sensitive to the size of supercell
when inducing a periodic strain gradient via first-principles
calculations.\cite{hong-jpcm10} The FCC obtained in the small
supercell in their work, taking into account only
the interactions between the local dipoles and strain
gradient, may not reflect the full contribution
to the FCC.\cite{yudin-unpub}

\begin{table*}
\caption{
Longitudinal FEC components at fixed $\E$ at room temperature.
$\mlone=\m\bl$, $\mltwo=\m\btr + 2 \m\btm$, and anisotropy
$\Delta=\mltwo -\mlone$. Units: nC/m. Note that the choice of units
is three orders or magnitude larger than in Table~\ref{tab:fixd1}. 
\mylabel{tab:fixe1}
}
\begin{ruledtabular}
\begin{tabular}{lrrrrrrrrrrrr}
 & & & 
\multicolumn{5}{c}{Even force}  &
\multicolumn{5}{c}{Mass-weighted force}  \\
\cline {4-8} \cline{9-13} 
& 
\multicolumn{1}{c}{$\mu\pel\blone$}  &
\multicolumn{1}{c}{$\mu\pel\bltwo$}  &
\multicolumn{1}{c}{$\mu\pld\blone$}  &
\multicolumn{1}{c}{$\mu\pld\bltwo$}  &
\multicolumn{1}{c}{$\mlone$}  &
\multicolumn{1}{c}{$\mltwo$}  &
\multicolumn{1}{c}{$\Delta$}  &
\multicolumn{1}{c}{$\mu\pld\blone$}  &
\multicolumn{1}{c}{$\mu\pld\bltwo$}  &
\multicolumn{1}{c}{$\mlone$}  &
\multicolumn{1}{c}{$\mltwo$}  &
\multicolumn{1}{c}{$\Delta$}  \\
\hline
C       & --1.0     & --0.9    &   0       &   0      & --1.0     &  --0.9   &0.1    &  0        & 0        & --1.0     & --0.9  &0.1      \\
Si      & --1.3     &--1.1     &     0     &   0      & --1.3     & --1.1    & 0.2   &  0        & 0        &  --1.3    &--1.1   &0.2     \\
MgO     &    --0.3  &   --0.5  &    --0.8  &--0.6     &   --1.1   &  --1.1   &  0.0  & --1.1     &  --0.9   &   --1.4   & --1.4  &  0.0  \\
NaCl    &    --0.1  &   --0.2  &    --0.3  &--0.2     &    --0.4  &    --0.4 &  0.0  &   --0.2   &--0.2     &    --0.3  &   --0.4&--0.1  \\
CsCl    &    --0.4  &   --0.4  &    --0.4  &--0.5     &    --0.6  &    --1.0 &--0.4  &   --0.5   &--0.8     &    --0.9  &   --1.2&--0.3  \\
BZO    &    --0.8  &   --1.0   &   --1.1   &--1.4     &   --1.9   &  --2.4   &--0.5  &  --2.0    &--2.2     &  --2.8    & --3.2  &--0.4 \\
BTO    &   --1.1   &   --1.4   & --333.2   &--435.4   & --334.3   & --436.8  &--102.5&--369.7    &--479.6   &--370.8    &--481.0 &--110.2\\
PTO    &   --1.5   &   --2.0   &--18.5     &--24.9    &  --20.0   &  --26.9  &--6.9  &--24.9     &--32.0    &  --26.4   & --34.0 &--7.6  \\
STO    &    --0.9  &   --1.0   &--36.0     &--42.0    &  --36.9   & --43.0   &--6.1  & --47.5    &--52.5    &  --48.4   &  --53.5 &--5.1  \\
\end{tabular}
\end{ruledtabular}
\end{table*}

We now turn to our results for the FECs at fixed $\E$.  According
to \equs{flexo-d2e}{flexo-frozen-d2e}, the total FEC $\m^\E$
and the electronic (frozen-ion) FEC $\m\pelE$ and are given by
multiplying $\m^D$ or $\m\pelD$ by the the static $\epsilon^0$
or optical $\epsilon^\infty$ respectively.  Because we are
interested in comparing with room-temperature experiments, we
used room-temperature static dielectric constants $\epsilon^0$
for all the materials as given in Table \ref{tab:properties1}.
Our results are given in Table \ref{tab:fixe1}, where $\mu\pld$
is obtained through $\mu\pld = \mu - \mu\pel$. From this table
we can see that the lattice contributions are larger  than
the electronic ones under fixed-$\E$ boundary conditions,
especially for the high-dielectric-constant materials.
It is also evident that the anisotropic flexoelectric response for
non-perovskite materials is small, indicating that the FxE tensor
of these materials is close that of an isotropic material having
only have two independent FEC components.\cite{quang-prsa11}
However, note that the experimental \BTO\ and \PTO\ materials
are tetragonal ferroelectrics at room temperature, and the
room-temperature $\epsilon^0$ values used in the conversion to
$\mu^\E$ for these materials were obtained from this tetragonal
structure.  Therefore, the results presented for \BTO\ and \PTO\
in Table \ref{tab:fixe1} are not fully consistent and should be
interpreted with caution.  We also assume for all materials
that the optical dielectric constants and fixed-$D$ FECs are
temperature-independent, but we expect this to be a rather
good approximation.

 From Table~\ref{tab:fixd1} it can be seen that the FECs at fixed
$D$ for the different perovskites are very similar, while instead
the fixed-$\E$ FECs reported in Table~\ref{tab:fixe1} show
dramatical variations. This
is because $\me$ is linearly scaled to the static dielectric
constant, as in Eq.~(\ref{eq:flexo-d2e}), and these perovskites
have very different dielectric constants as shown in
Table~\ref{tab:properties1}. \BTO\ has the largest FEC due to its
large dielectric constant, which also explains the large
FEC observed in the work of
Ma and Cross;\cite{ma-apl01,ma-apl01-1,ma-apl02,ma-apl03,ma-apl05,ma-apl06}
all of the materials they measured
displayed large dielectric constants (typically several thousands).
However, they also observed that the FEC does not exactly scale
linearly with the 
dielectric constants in some materials, which may indicate that 
$\mu\pD$ also has some weak temperature dependence.
Their measurements may also be affected by surface effects not
considered here.

Our flexoelectricity theory is valid for materials with any symmetry, but
our current first-principles calculations are limited to cubic
materials.  However, our previous work has shown that ferroelectric
tetragonal \BTO\ has a similar $\m_{1111}$ value
(along the [001] direction) as for cubic \BTO,\cite{hong-jpcm10}
and ferroelectric tetragonal \PTO\ has a similar
$\mu\pel_{1111}$ (along [001]) as for cubic \PTO.\cite{hong-prb11}
These results hint that $\mu_{1111}$ may have similar values
in tetragonal and cubic phases of a given perovskite material.

\subsection{Transverse and full Cartesian flexoelectric response}
\mylabel{sec:apxm}

In order to make closer contact with experiment, in this section our
goal is to present results for the fixed-$\E$ FECs in the Cartesian
frame.  If $\mlone$, $\mltwo$, and $\mtr$ are known, it is straightforward
to invert \equr{mlone}{mtr} to obtain $\m\bl$, $\m\btr$, and $\m\btm$.
However, as discussed in Sec~\ref{sec:flexo-part} and more fully in
Appendix~\ref{app:current}, we can only obtain the longitudinal
contributions $\mu\blone$ and $\mu\bltwo$ from our
first-principles calculations based on the charge-response formulation;
additional terms from the current-response formulation would be needed
to obtain $\mtr$.  Since $\m\bl=\mlone$, there is no ambiguity about
its value, but $\m\btr$ and $\m\btm$ cannot be obtained individually
without access to $\mtr$.

Another way to see the problem is that we can imagine obtaining
$\mu\pld_{1122}$  or $\mu\pld_{1221}$ at fixed $D$ directly from
\equ{muldT}.  However, this requires a knowledge of $T^D_{1122}$,
which is not one of the raw ingredients available to us in
Table~\ref{tab:rawdata}.  Instead, we have $T^\E_{1122}$, but
converting this to $T^D_{1122}$ would require the use of
\equ{K2d-e}; this in turn requires $\m\pelD_{1122}$, which is
not available without the current-response calculation.

So the problem is that we do not have the
lattice quadrupole contribution $\mu\plqJ$ to $\m\plqT$
in \equs{mulq}{mulqj}, nor do we have the electronic contribution
$\mu\pelJ=\m\pelT$ in \equs{muel}{muelj}.
The lattice quadrupole contribution $\m\plq$ vanishes by symmetry
for all of our compounds except for C and Si, where it
makes a contribution only to the transverse component $\mtr$.
We have computed these contributions, neglecting any current-response
contribution  $\mu\plqJ$, following the discussion in
Sec.~\ref{sec:modes}. We find  $\gamma = 0.14$ (0.72)\,Bohr,
$q= -0.10$ (0.12)\,e\,Bohr, and
$\mu\plqT = -0.84$\,pC/m ($1.5$\,pC/m) for C (Si) respectively.
These values are much smaller than the other values in
Table~\ref{tab:fixd1}, and we speculate that the current-response
contributions are probably small too.  We do not consider the
lattice quadrupole contributions any further here.

For the materials other than C and Si, the only missing ingredient is
the electronic contribution $\mu\pelT$
to the transverse response, which comes
entirely from the current-response
contribution $\mu\pelJ$ in \equ{muelj}. We have essentially
no information about this contribution.  On the other hand,
we can still calculate the full lattice-dipole contribution
$\mu\pld$, since there is no current-response correction for
this term.  To do so, we move one layer of atoms along the $y$
direction under fixed-$\E_y$ boundary conditions and
obtain $T\pE_{1122}$ and thus
$\mu\pldE_{1122}$  and $\mu\pldE_{1221}$.\cite{notes-Te}
For high-$K$ materials
in which the lattice contribution dominates the FEC, calculating
$\mu\pld$ is a good approximation for the total FEC response.
However, for the low-$K$
materials in which the electronic contribution $\mu\pel$ is
comparable to the lattice contribution, we need to obtain
$\mu\pel _{1122}$ and $\mu\pel _{1221}$ for the full FEC tensor.
Here, we introduce the assumption that $\mu\pel _{1122} =
\mu\pel _{1221}$, i.e. $\mu\pelT =\mu\pelJ= 0$, to obtain a rough
first approximation to the full FEC tensor.  A proper calculation
would require the inclusion of the current-response terms of
Appendix~\ref{app:current}, which are not currently implemented
in this work.

\begin{table*}
\caption{
Cartesian FECs at fixed $\E$ using the assumption 
$\mu\pel _{1122}=\mu\pel _{1221}$.
Room-temperature $\epsilon^0$ values are used for perovskites
(see Table \ref{tab:properties1}).
First column is experimental Poisson's ratio $t$ (dimensionless);
others are FECs in nC/m.  FECs $\m$ and $g$ are defined in terms of
unsymmetrized and symmetrized strains respectively and are related
by \equr{gma}{gmc}.
\mylabel{tab:full-FEC}
}
\begin{ruledtabular}
\begin{tabular}{ldrrrrrrrrrr}
&
\multicolumn{1}{c}{Poisson's} &
\multicolumn{5}{c}{Even force}  &
\multicolumn{5}{c}{Mass-weighted force}  \\
\cline {3-7} \cline{8-12}
&
\multicolumn{1}{c}{ratio} &
\multicolumn{1}{c}{$\m_{1111}$}  &
\multicolumn{1}{c}{$\m_{1122}$}  &
\multicolumn{1}{c}{$\m_{1221}$}  &  & &
\multicolumn{1}{c}{$\m_{1111}$} &
\multicolumn{1}{c}{$\m_{1122}$} &
\multicolumn{1}{c}{$\m_{1221}$} \\
&
\multicolumn{1}{c}{$t$}  &
\multicolumn{1}{c}{$g_{1111}$}  &
\multicolumn{1}{c}{$g_{1221}$}  & &
\multicolumn{1}{c}{$g_{1122}$}  &
\multicolumn{1}{c}{$g^{\rm eff}$}  &
\multicolumn{1}{c}{$g_{1111}$}  &
\multicolumn{1}{c}{$g_{1221}$}  & &
\multicolumn{1}{c}{$g_{1122}$} &
\multicolumn{1}{c}{$g^{\rm eff}$}  \\
\hline
C &    0.1\tablenotemark[1]
 & --1.0    & --0.3    &  --0.3   &   --0.3  &--0.2      & --1.0     & --0.3   &--0.3       & --0.3      & --0.2    \\
Si &   0.22\tablenotemark[1]
 &--1.3     & --0.4    & --0.4    & --0.4    &   0.0     &--1.3      &--0.4    & --0.4      & --0.4      &   0.0    \\
MgO &  0.18\tablenotemark[1]
 &--1.1     &   --0.3  &  --0.4   & --0.6    &--0.3      &--1.4      &   --0.4 &   --0.5    & --0.7      & --0.3    \\
NaCl & 0.25\tablenotemark[1]
 &--0.2     &     0.0  &   --0.2  & --0.4    &--0.2      & --0.2     &    0.0  &   --0.2    & --0.4      &--0.3     \\
CsCl & 0.27\tablenotemark[1]
 &    --0.8 &    --0.2 &    --0.4 &    --0.6 & --0.2     &   --0.9   &    --0.5&    --0.3   &   --0.2    & 0.0      \\
BZO &  0.24\tablenotemark[1]
 &--1.9     &      0.0 & --1.2    &  --2.3   & --1.3     & --2.8     & --0.2   &  --1.5     &   --2.7    & --1.4      \\
BTO &  0.27\tablenotemark[1]
 & --334.3  &   13.8   &--225.3   & --464.3  &--248.7    & --370.8   &   --2.5 &--239.2     & --475.9    &--247.3    \\
PTO &  0.31\tablenotemark[1]
 & --20.0   &     0.1  &--13.5    & --27.2   &--12.4     &  --26.4   &   --2.0 & --16.0     &  --30.0    & --12.3     \\
STO &  0.24\tablenotemark[1]
 & --36.9   &  --1.4   &--20.8    & --40.2   & --21.6    &  --48.4   &   --5.0 & --24.2     &  --43.5    & --21.4     \\
\end{tabular}
\tablenotetext[1]{Experimental values:
 C, Ref.~\onlinecite{c};
 Si, Ref.~\onlinecite{si};
 MgO, Ref.~\onlinecite{mgo};
 NaCl, Ref.~\onlinecite{nacl};
 CsCl, Ref.~\onlinecite{cscl};
 BZO, Ref.~\onlinecite{bzo};
 BTO, Ref.~\onlinecite{bto};
 PTO, Ref.~\onlinecite{pto};
 STO, Ref.~\onlinecite{sto}.
}
\end{ruledtabular}
\end{table*}

The results for the full fixed-$\E$ FEC tensor at room temperature,
making use of this assumption, is presented in
Table~\ref{tab:full-FEC}.  We present the results in terms of
the FEC components defined in terms of both {\em unsymmetrized} strains
($\m$) and {\em symmetrized} strains ($g$).  We also give
values for the effective beam-bending FEC defined
as~\cite{zubko-prl07,zubko-prl07-err} 
\beq
g^{\rm eff} = -t g_{1111} + (1-t) g_{1122} \;,
\mylabel{eq:eff-FEC}
\eeq
where $t=C_{1122}/(C_{1111}+C_{1122})$ is Poisson's ratio with
$C$ being the elastic stiffness tensor.  Experimental values of
$g^{\rm eff}$ were reported in Table~\ref{tab:full-FEC} for our
materials of interest.  Letting $x$ and $z$ be the thickness and
length directions respectively, Eq.~(\ref{eq:eff-FEC}) follows
from the the expression for the displacement field in the case of
pure beam bending, given by\cite{landau-book70} $ u_x = -[z^2 +
t (x^2 - y^2)]/2R$, $u_y = -t x y /R$, and $u_z = x z /R$, where
$R$ is the radius of curvature of the neutral surface of the
beam. Thus the symmetrized strain gradient is $\nu^s_{xxx}
= \nu^s_{yyx} = - t \,\nu^s_{zzx}$, leading to
Eq.~(\ref{eq:eff-FEC}).

As we discussed in Sec.~\ref{sec:relaxed-ion}, the lattice
contribution to the FECs has a dependence on the choice of force pattern
because of the force density that arises due to the non-zero divergence
of the stress in a general strain-gradient configuration.
However, in the beam bending experiment, the beam is in
local static equilibrium at every interior point, so there is no
such force density; the contributions coming from individual elements
of the strain gradient tensor cancel each other.
Therefore, the effective beam-bending FEC $g^{\rm eff}$ should not
depend on the choice of force pattern in our calculation.  This
can be confirmed by referring to Table~\ref{tab:full-FEC}, where
$g^{\rm eff}$ obtained from different force patterns are
indeed the same to within the numerical precision of the calculations.
This confirms that although the individual FEC components are
force-pattern dependent coming from the first-principles calculations,
using them for problems of static equilibrium should not cause any
problem.  Conversely, it follows that it may be problematic to
measure the full set of FEC components experimentally from
static-equilibrium experiments,
since only special linear combinations of strain-gradient
components, for which the force density vanishes, are accessible
in this way.

Since quantities related to elastic properties are usually defined
in the literature for symmetrized strain, we focus on the
FECs $g\babcd$ related to symmetrized strain for the remainder
of this subsection. From
Table~\ref{tab:full-FEC}, we can see that the values of
$g_{1111}$ and $g_{1122}$ are comparable, but the
value of $g_{1221}$ is
much smaller than the other two components,
indicating that the gradient of the shear strain
$\epsilon_{21,2}$ makes only a very weak contribution to the
flexoelectric response (under the assumption $\mu\pel=0$), in
agreement with the results of
Ponomareva \textit{et al.}\cite{ponomareva-prb12}

According to Eq.~(\ref{eq:eff-FEC}), in order to obtain a large
bending flexoelectric response, a large $g_{1122}$ and a small
$g_{1111}$ are preferred, as well as a small Poisson's ratio.  Even
if the individual FEC components are all negative, the effective
beam-bending FEC can still be positive, depending on the ratio of
$g_{1111}$ to $g_{1122}$. Specifically, $g^{\rm eff}$ is has a
positive value if $g_{1111} / g_{1122} > (1-t) / t $ (assuming
$g_{1111}$ and $g_{1122}$ are negative).

\subsection{Comparison with experiment and previous theoretical
results}

As mentioned in the Introduction, large discrepancies between
experimental and theoretical results for the FECs have been reported.
For example, for high-$K$ materials, the
experimental effective FECs are usually reported to be on the order of
$\mu$C/m\cite{ma-apl01,ma-apl01-1,ma-apl02,ma-apl03,ma-apl05,ma-apl06}
with a positive sign, while the theoretical results are typically on the
order of nC/m with negative sign.\cite{maranganti-prb09,hong-jpcm10}
One possible reason is that the theoretical results
are calculated at 0\,K while the experimental results are measured
at room temperature (or above the Curie temperature), and very
strong dependence of the static
dielectric constant can contribute to this large
discrepancy. Regarding the sign problem, the effective
beam-bending coefficient is given by Eq.~(\ref{eq:eff-FEC}), and
as discussed in the previous subsection, the
negative individual components can give a positive effective
beam-bending FEC  even if $g_{1111}$ and $g_{1122}$ are
negative.~\cite{notes-gpos}

As we showed in the previous section, the individual FEC components are
dependent on the force pattern adopted for the first-principles
calculations.  Therefore, we should not expect that individual
FEC tensor components, such at $\mu_{1111}$, can be compared
directly between theory and experiment.

In order to obtain the full FEC tensor, we need to have
information about $\mu_{\rm T}$. In Sec.~\ref{sec:apxm} we assumed
that $\mu\pelT = 0$ in order to obtain a rough estimate of the
the full FEC tensor as presented in
Table~\ref{tab:full-FEC}. However, it can be seen that the
effective beam-bending FEC computed in that way
does not agree with available experiment
results. For example, $g^{\rm eff}$ has been reported to be
6.1\,nC/m for \STO\ single crystals\cite{zubko-prl07,zubko-prl07-err}
and $9\,\mu$C/m for \BTO\ ceramics\cite{ma-apl06} at room temperature,
while we obtain $-$22\,nC/m and $-0.246\,\mu$C/m for
\STO\ and \BTO\ respectively.
There are several possible reasons for this discrepancy.
First, it may well be that our assumption that $\mu\pelT = 0$,
introduced to obtain the full FEC tensor in Table~\ref{tab:full-FEC},
is strongly inadequate.\cite{notes-gpos}
Second, the experiment results include surface
effects that have not been included here.
To do so would require computing the surface contributions as discussed
in Ref.~[\onlinecite{hong-prb11}], which is beyond the scope of the
present paper.

Hong \textit{et al.}\cite{hong-jpcm10} also calculated the longitudinal FECs
for \STO\ and \BTO\ at fixed $D$, including electronic and lattice
contributions, by imposing a strain wave of cosine form for the
A atoms in the supercell and relaxing all other atoms.  This corresponds
to a choice of force pattern in which all of the force is on the
A atoms.
Their results are
$\md_{1111} = -0.37 \pm 0.03$ nC/m for \BTO and
$\md_{1111} = -1.38 \pm 0.65$ nC/m for \STO.  It can be
seen that our results agree with their \BTO\ result in order
of magnitude and sign, with the remaining
discrepancy coming mostly from the difference in force patterns.

\section{Conclusion}
\mylabel{sec:conclusion}

A general and unified first-principles theory of piezoelectric
and flexoelectric tensors has been developed and presented here.
The longitudinal contributions to the flexoelectric tensor can
be computed from the dipoles associated with
strain-gradient-induced displacements (lattice dipole), quadrupoles
associated with strain-induced displacements (lattice quadrupole),
and octupoles associated with an ideal strain gradient (electronic).
The full tensor also requires the transverse part, which has
contributions that can only be obtained from the adiabatic currents
that flow in response to the flexoelectric displacements.

While the full formalism is presented in Appendix~\ref{app:current},
we have implemented only the charge-response formalism in the present
work, following the equations presented in the main text and working
in the framework of first-principles density-functional calculations.
We have paid careful attention to the distinction
between FECs computed at fixed $\E$ vs.\ fixed $D$ and presented
the relationships connecting them.  We have argued that the
FECs at fixed $D$ provide a characteristic ``ground-state bulk
property'' that can be used to predict finite-temperature
fixed-$\E$ properties by scaling to the dielectric constant.
We also show how the FCCs can be computed from our approach
and used in a similar way as for the fixed-$D$ FECs.

A practical supercell-based method is proposed to calculate the FECs
from first principles and is demonstrated by computing the
coefficients of several cubic insulating materials, namely C, Si, MgO,
NaCl, CsCl, BaZrO$_3$, BaTiO$_3$, PbTiO$_3$ and SrTiO$_3$. It is
found that the FECs at fixed $D$ are on the order of $-0.1$\,nC/m for
all these materials, and their FCCs are in the range of $-10$ to $-20$\,V,
indicating that when large FECs are found in experiment, it is
likely to arise from a large dielectric constant, or possibly
from surface effects not treated explicitly here.
Therefore, searching for large dielectric-constant
materials is a good way to obtain materials with a large FxE response.
The FECs computed from our first-principles theory at fixed $\E$
still do not agree well with available experiment results,
even after considering the relations between different electric
boundary conditions. However, this discrepancy is two orders of
magnitude less severe that some previous discrepancies 
between theory and experiment for the FECs (i.e., for \BTO).
When surface effects
are treated properly, it is hoped that this discrepancy will
become even smaller.  

Our calculations show that the lattice contribution to the FECs
depends on the force pattern applied in the unit cell to maintain
the strain gradient. Therefore, the total FECs are also dependent
on  the force pattern, and it is not meaningful to compare with
FECs computed using a different force pattern. However, for a
system in static equilibrium (zero stress gradient) the force-pattern
dependence should cancel out, and we have confirmed that this
is indeed the case from our numerical calculations.

The full FEC tensor is critical for an understanding of the FxE
response in cases of complicated strain distributions, as well as
for the design of functional FxE devices.  In general it is
impossible to obtain all the elements of this tensor using only
the charge-response formulation, even for cubic materials which
have only three independent components.   Therefore it
is clearly of interest to develop a full implementation based
on the current-response tensors as described in Appendix~\ref{app:current}.
The implementation of such a method, and the calculation of the
remaining terms in the FxE response, remains as an important
avenue for future work.

\acknowledgments

This work was supported by ONR Grant N00014-12-1-1035. Computations were
performed at the Center for Piezoelectrics by Design. We would
like to thank R. Resta, K.M.~Rabe, D.R.~Hamann, and M.~Stengel for
useful discussions.

\appendix

\section{Current-response formalism and transverse tensor components}
\mylabel{app:current}

In Sec.~\ref{sec:charge-response}, we carried out a derivation
that expresses the flexoelectric tensor in terms of changes in
the charge density induced by the atomic displacements associated
with the strain gradient. However, this procedure only determines
a part of the flexoelectric response, which we denote as the
``longitudinal'' part since it corresponds to the the longitudinal part
of the current-density field that arises as the deformation of
the material is adiabatically turned on.  In this Appendix, we
derive the full expression in terms of the current-density response,
thus clarifying the status of the expressions given in
Sec.~\ref{sec:charge-response}.

As we discussed in our previous work,\cite{hong-jpcm10} the choice of
the induced current density instead of the induced charge density provides
a more complete description of the response.  We define
\beq
\PP\biab(\rr-\R_{lI})=\frac{\pa \mathcal{J}\ba (\rr)}{\pa \dot{u}_{lI,\b}}
\mylabel{eq:f2}
\eeq
to be the current density $\mathcal{J}\ba(\rr)$ in Cartesian
direction $\a$ resulting from the adiabatic motion of atom $I$
in cell $l$ at some small velocity $\dot{u}_{lI,\b}$ along the
$\b$ direction, keeping
all other atoms fixed.  We can simply think of this as the local
polarization field $\Delta \P(\rr)$ induced by the displacement
of the atom.  Such a quantity is not generally well-defined
for a \textit{finite} adiabatic deformation; while its cell
average is fixed by the change in Berry-phase polarization and
its longitudinal (curl-free) part is fixed by the change in
ground-state charge density, its transverse (i.e., divergence-free)
part is not guaranteed to be independent of path.  However, for an
\textit{infinitesimal} displacement, as arises here in the linear
response to a small strain gradient, there is no such ambiguity.

We then define the moments of the induced current densities in
analogy with Eqs.~(\ref{eq:Q1}-\ref{eq:Q3}) as
\bea
&&
J\pzer _{I,\a\b}=\int d\rr\, \,\PP\biab(\rr)\, ,
\mylabel{eq:J0} \\
&&
J\pone _{I,\a\b\c}=\int d\rr\, \PP\biab(\rr)\, r\bc \, ,
\mylabel{eq:J1} \\
&&
J\ptwo _{I,\a\b\c\d}=\int d\rr\, \PP\biab(\rr)\, r\bc\, r\bd \,  .
\mylabel{eq:J2}
\eea
Note that $J\ptwo _{I,\a\b\c\d}$ is symmetric in indices
$\c\d$, but otherwise these tensors are general.  The
charge-response tensors of Eqs.~(\ref{eq:Q1}-\ref{eq:Q3}) are
related to the current-response tensors via
\bea
&&\!\!\!
Q\pone _{I,\a\b}=J\pzer _{I,\a\b} \;,
\mylabel{eq:JQ1} \\
&&\!\!\!
Q\ptwo _{I,\a\b\c}=J\pone _{I,\a\b\c}+J\pone _{I,\c\b\a} \;,
\mylabel{eq:JQ2} \\
&&\!\!\!
Q\pthr _{I,\a\b\c\d}=
J\ptwo _{I,\a\b\c\d}+J\ptwo _{I,\c\b\a\d}+J\ptwo _{I,\d\b\a\c} \;.
\mylabel{eq:JQ3}
\eea
These equations follow after integration by parts using the Poisson continuity
condition
\beq
f\bib(\rr)=-  \partial\ba \PP\biab(\rr) \;.
\mylabel{eq:poisson}
\eeq
For example, inserting \equ{poisson} in \equ{Q2} gives
\bea
Q\ptwo _{I,\ \a\b\c} &=& -\int d\rr\, r\ba r\bc \,\partial_\mu \PP_{I,\mu\b}(\rr)
\nn
&=& \int d\rr\, \PP_{I,\mu\b}(\rr) \,  \partial_\mu(r\ba r\bc)
\eea
which leads to \equ{JQ2} using \equ{J1} after noting that
$\partial_\mu(r\ba r\bc)=\delta_{\mu\a}r\bc+\delta_{\mu\c}r\ba$.

Eqs.~(\ref{eq:JQ1}-\ref{eq:JQ3}) make it clear that the
charge-response and current-response tensors are closely related.
In fact, $Q\pone$ and $J\pzer$ are identical.  For the higher-order
tensors, however, the charge-response tensors contain less
information.  Essentially, they only contain information about
the longitudinal part of the induced current response and lack the
additional information contained in the transverse part.  A simple
way to see this is just to count elements: recalling the symmetries of
the various tensors under interchanges of indices, we note that
$Q\ptwo _{I,\a\b\c}$ has 3$\times$6=18 independent elements
while $J\pone _{I,\a\b\c}$ has 3$\times$3$\times$3=27, and
$Q\pthr _{I,\a\b\c\d}$ has 3$\times$10=30 elements while $J\ptwo
_{I,\a\b\c\d}$ has 3$\times$3$\times$6=54.  This makes it clear
that some information is missing from the charge-response tensors.

To make this more precise, we define the longitudinal parts of the
$J$ tensors to be
\beq
J\poneL_{I,\a\b\c}=\half\left(J\pone_{I,\a\b\c}+J\pone_{I,\c\b\a}\right)
\eeq
and
\beq
J\ptwoL_{I,\a\b\c\d}= \frac{\textstyle{1}}{\textstyle{3}}
\left( J\ptwo_{I,\a\b\c\d} + J\ptwo_{I,\c\b\a\d} + J\ptwo_{I,\d\b\a\c} \right)
\eeq
and the transverse parts to be the remainders
\bea
&& J\poneT_{I,\a\b\c}=J\pone_{I,\a\b\c}-J\poneL_{I,\a\b\c} \;, \\
&& J\ptwoT_{I,\a\b\c\d}=J\ptwo_{I,\a\b\c\d}-J\ptwoL_{I,\a\b\c\d} \;.
\eea
The longitudinal current-response tensors describe
the moments of the curl-free part of $\P_{I,\a}$ and can
be written, using Eqs.~(\ref{eq:JQ2}-\ref{eq:JQ3}), as
\bea
&& J\pzer_{I,\a\b}= Q\pone _{I,\a\b} \;,
\mylabel{eq:JQL1} \\
&& J\poneL_{I,\a\b\c}=\half Q\ptwo _{I,\a\b\c} \;,
\mylabel{eq:JQL2} \\
&& J\ptwoL_{I,\a\b\c\d}=\frac{\textstyle{1}}{\textstyle{3}} Q\pthr _{I,\a\b\c\d}
\mylabel{eq:JQL3}
\;,
\eea
thus reproducing the information in the charge-response tensors.  On the
other hand, the transverse current-response tensors $J\poneT$ and
$J\ptwoT$ contain new information that is not otherwise available.

The entire derivation of Sec.~\ref{sec:charge-response} can now be
repeated using the current-response formalism.  \equ{rhof} is replaced by
\beq
P\ba(\rr)= \sum_{lI\t} \PP_{I,\a\t}(\rr - \R_{lI}) \, u_{lI,\t}
\mylabel{eq:PP}
\eeq
and its Fourier transform is
\beq
P\ba(\k) = \Vc^{-1} \sum_{I\t} W_{I\t\b}(\k) \, \PP_{I,\a\t}(\k) \, u_{0\b}
\mylabel{eq:Pk}
\eeq
where $\PP_{I,\a\t}(\k)$ is the Fourier transform of $\PP_{I,\a\t}(\rr)$.
\equ{expan} is then replaced by
\begin{widetext}
\bea
P\ba(\k) &=& \Vc^{-1}\sum_I \left[
  i\,\sum_\t J\pzer_{I\a\t} \G_{I\t\b\c} \, k_\c
  -\, i\,  J\pone_{I\a\b\nu} k_\nu
\right] u_{0\b} \nn
&& +\Vc^{-1}\sum_I \left[
  - \sum_\t J\pzer_{I\a\t} N_{I\t\b\c\d} \, k_\c k_\d
  +\sum_\t J\pone_{I\a\t\nu} \G_{I\t\b\c} k_\nu k_\c
  -\frac{1}{2} J\ptwo_{I\a\b\nu\mu} k_\nu k_\mu
\right] u_{0\b} \nn
&& +\ldots
\mylabel{eq:expanP}
\eea
\end{widetext}

Comparing this with \equ{Peu} we conclude that
\beq
e_{\a\b\c} =  \Vc^{-1} \sum_{I\t} J\pzer_{I\a\tau} \G_{I\tau\b\c} \,
 - \Vc^{-1} \sum_I J\pone_{I\a\b\c}
\;.
\mylabel{eq:e-curr}
\eeq
Using Eqs.~(\ref{eq:JQL1}-\ref{eq:JQL2})
we find that the lattice-dipole part $e\pld$ is still given by
\equ{eld} while the electronic part $e\pel$ becomes
\beq
e\pel\babc= \Vc^{-1} \sum_I \left( -\half Q\ptwo\biabc- J\poneT_{I\a\b\c}
\right) \;.
\eeq
Thus, the free contribution $e\pel\babc$ in \equ{eel} has now been determined
and can be seen to represent precisely the transverse components that
were omitted in the charge-response derivation.

In a similar way, we can now obtain the full flexoelectric
tensor.  Remembering that $\mu\babcd$ is forced by definition
to be symmetric in the last two indices, we find that the the
contributions to the flexoelectric tensor, \equ{mu-three}, are
now given by
\bea
\mu\babcd\pld &=& \Vc^{-1} \sum_{I\t} J\pzer\biat\,N\bitbcd
\;, \\
\mu\babcd\plq &=& -\half \Vc^{-1} \sum_{I\t} \left(
   J\pone\biabc\,\G\bitbd \right. \nn
&& \hspace{1.8cm}  \left. +  J\pone\biabd\,\G\bitbc \right)
\;, \\
\mu\babcd\pel &=& \half \Vc^{-1} \sum_I J\ptwo\biabcd
\;.
\eea
Once again, Eqs.~(\ref{eq:mulq}-\ref{eq:muel})
are recovered, but
now we can identify the missing transverse pieces as
\bea
\mu\babcd\plqJ &=& -\half \Vc^{-1} \sum_{I\t} \left(
   J\poneT\biabc\,\G\bitbd \right. \nn
&& \hspace{1.8cm}  \left. +  J\poneT\biabd\,\G\bitbc \right) \;,
\mylabel{eq:mulqj} \\
\mu\babcd\pelJ &=&
  \half \Vc^{-1} \sum_I J\ptwoT\biabcd \;.
\mylabel{eq:muelj}
\eea
This completes the full derivation of the flexoelectric response
tensor using the current-response formalism.

Methods for computing the transverse parts of the current-response
tensors $J\poneT$ and $J\ptwoT$ have not been developed as part of
the present work.
No extra contributions are needed for the lattice dipole
contribution, and the lattice quadrupole terms vanish for
all of the cubic materials considered in this work except for C
and Si.  For the
electronic contribution, however, we are only able to report
on the longitudinal contributions  $\m^{\rm e\hspace{0.2pt}l,L1}$ 
and $\m^{\rm e\hspace{0.2pt}l,L2}$, leaving the calculation of
 $\m^{\rm e\hspace{0.2pt}l,T}$  for future work.

\section{Pseudo inverse of force constant matrix}
\mylabel{app:inverse}

We begin by restating the problem posed in Sec.~\ref{sec:force-pattern},
simplifying the notation by dropping the Cartesian indices.
This is clearly sufficient for the binary cubic materials
considered here, since the force-constant matrix is block-diagonal
in the Cartesian representation, and the procedure outlined applies
to each $N\times N$ block ($N$ is the number of atoms per cell).
(For the perovskites, the transformation to symmetry mode variables
outlined in Appendix~\ref{app:oxygen} is performed first.
For more complex crystals, the force-constant matrix would
be block-diagonalized by IR-active irrep before the procedure
would be applied, with $N$ replaced by the number of copies of the
irrep.)

With this simplification, the problem is as follows.  We are given
a force-constant matrix $K=K^T$ obeying the acoustic sum rule
$\sum_j K_{ij}t_j=0$, where $t_j$ is a vector all of whose elements
are 1, and a set of
weights $w_i$ specifying a ``force pattern.''  We wish to construct
a pseudo-inverse $\Jw$ having the property that \equ{fbalance} is
obeyed, i.e.,
\beq
f\pext_i-\sum_{jk} K_{ik}\,\Jw_{kj}\,f\pext_j=(\sum_j t_j f\pext_j)\,w_i
\eeq
for any external force vector $f\pext$.
To simplify the notation we use a bra-ket notation for vectors with
implied matrix-matrix and matrix-vector products, so that this is
equivalent to
\beq
K\,\Jw=I-\ket{w}\bra{t}
\mylabel{eq:targ}
\eeq
with $\ip{w}{t}=1$.

The construction proceeds as follows.  Construct an $N\times N$
matrix $E$ whose first column is $\ket{w}$ and whose remaining
columns are all orthogonal to $\ket{t}$, being sure to keep the
columns linearly independent.  Also define $D=(E^T)^{-1}$, i.e.,
the matrix whose columns are the duals to those of $E$
(that is, $D^TE=I$).  This means that the first column of $D$ is
just $\ket{t}$. We can think of $D$ and $E$ as giving
the transformations back and forth between the original atomic
displacements and a set of mode variables of which the first
is the uniform translation.

Next let $D_r$ and $E_r$ be the $N\times(N-1)$ rectangular
matrices constructed by dropping the first column of $D$ and
$E$ respectively.  Letting $\ket{o}$ be the vector $(1,0,\ldots)$,
this can be written as $D_r=D-\ket{t}\bra{o}$ and
$E_r=E-\ket{w}\bra{o}$, and it follows that
\beq
E_r D_r^T = I -\ket{w}\bra{t}
\mylabel{eq:deficit}
\eeq
after using that $D\ket{o}=\ket{t}$ and $E\ket{o}=\ket{w}$.
Since $K$ obeys the acoustic sum rule, the first row and
column of $D^T K D$ are zero, and $K$ is fully represented
by the ``reduced'' matrix
\beq
K_r=D_r^T K D_r \quad \Leftrightarrow \quad
K=E_r K_r E_r^T \;.
\eeq
Then our solution is to set
\beq
\Jw= D_r (K_r^{-1}) D_r^T
\eeq
which is well-defined because the reduced matrix $K_r$ is
non-singular.  Substituting into \equ{targ} we get
\bea
K \Jw &=& (E_r K_r E_r^T)(D_r K_r^{-1}D_r^T) \nn
      &=& E_r K_r K_r^{-1}D_r^T \nn
      &=& E_r D_r^T = I -\ket{w}\bra{t} \;,
\eea
where \equ{deficit} was used on the last line.  This
satisfies \equ{targ}, showing that $\Jw$ is indeed the needed
pseudo-inverse.

\section{Oxygen in the perovskites}
\mylabel{app:oxygen}

As discussed in the main text, the site symmetry of individual O
atoms in cubic perovskites is not cubic, and some of the
space-group operations interchange O atoms.  To handle this
case, it is convenient to introduce ``mode coordinates.''

We define O1, O2 and O3 as the
oxygen atoms displaced by $a/2$ from the central Ti along
$\hat{x}$, $\hat{y}$, and $\hat{z}$, respectively.
Taking SrTiO$_3$ as our example system,
we start by considering zone-center phonons and carrying out a linear
transformation between the 15 sublattice displacement variables
$u_{I\t}$ describing the displacement of sublattice
$I$\,=\,\{Sr,\,Ti,\,O1,\,O2,\,O3\}
in Cartesian direction $\t$\,=\,$\{x,y,z\}$, and symmetrized
mode variables $\xi_{\sigma\t}$ that we choose to define as
\bea
&&
\xi_{1x} = u_{\Sr x}\,, \qquad
\xi_{1y} = u_{\Sr y}\,, \qquad
\xi_{1z} = u_{\Sr z}\,, \nn
&&
\xi_{2x} = u_{\Ti x}\,, \qquad
\xi_{2y} = u_{\Ti y}\,, \qquad
\xi_{2z} = u_{\Ti z}\,, \nn
&&
\xi_{3x} = u_{\O 1 x}\,, \qquad
\!\xi_{3y} = u_{\O 2 y}\,, \qquad
\!\!\xi_{3z} = u_{\O 3 z}\,, \nn
&&
\xi_{\{4,5\}x} = (u_{\O 3 x} \pm u_{\O 2 x})/\sqrt{2}\,, \nn
&&
\xi_{\{4,5\}y} = (u_{\O 1 y} \pm u_{\O 3 y})/\sqrt{2}\,, \nn
&&
\xi_{\{4,5\}z} = (u_{\O 2 z} \pm u_{\O 1 z})/\sqrt{2}\,,
\mylabel{eq:A}
\eea
where $\{4,5\}$ means that the plus and minus apply to
case 4 and 5 respectively.
Here $\sigma$ is a label running over $\sigma={1,2,3,4}$ for the
four copies of the IR-active $\GOF$ irrep,
while $\sigma=5$ corresponds to the IR-silent $\GTF$ irrep.
We can summarize this as
\beq
\xi_{\sigma\t}=\sum_{I\t'} A_{\sigma\t,I\t'} \, u_{I\t'}
\eeq
where the elements of $A_{\sigma\t,I\t'}$ are given in \equ{A}.

We have chosen an orthogonal transformation, $A^{-1}=A^T$,
so that forces transform in the same way,
\beq
\ft_{\sigma\t} = \sum_{I\t'} A_{\sigma\t,I\t'} \, f_{I\t'} \;.
\mylabel{eq:ftran}
\eeq
The $T$ tensor elements of \equ{Tdef} will also transform in
the same way,
\beq
\Tt_{\sigma\t,\b\c\d} = \sum_{I\t'} A_{\sigma\t,I\t'} \, T_{I\t'\b\c\d}
\;,
\eeq
which is essentially \equ{Ttrans} using $A^{-1}=A^T$.

\subsection{Original frame}

\begin{figure}
\begin{center}
\includegraphics[width=3.0in]{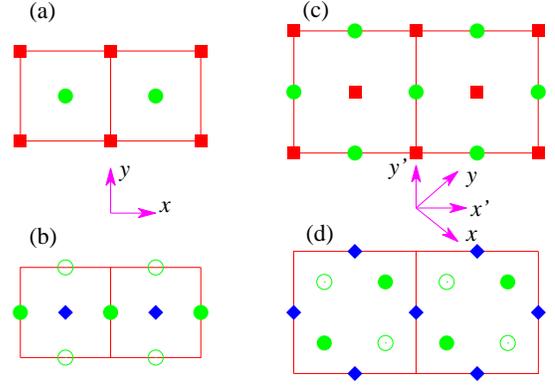}
\end{center}
\caption{\mylabel{fig:frame}
(Color online) ABO$_3$ perovskite atomic geometry in (a-b) original
Cartesian frame, and (c-d) 45$^\circ$ rotated frame, as appropriate
to the two supercells of Fig.~\ref{fig:model} respectively.
(a) and (c): slice at $z=0$;
filled squares (red) are A and
filled circles (green) are O$_3$ atoms.
(b) and (d): slice at $z=c/2$;
filled diamonds (blue) are B,
filled circles (green) are O$_1$, and
open circles (green) are O$_2$ atoms.
}
\end{figure}

Our main interest here is the calculation of these $T$ tensor
elements.  We start with the original supercell extended along
$x$ as in Fig.~\ref{fig:model}(a) and detailed in
Fig.~\ref{fig:frame}(a-b), and consider the forces in response
to a longitudinal strain gradient $\nu_{xxx}$.  In this case all
forces are along $x$ by symmetry, and from \equ{ftran} with $A$
given by \equ{A} it follows that
\bea
\ft_{3x} &=& f_{\O1x}\;, \nn
\ft_{4x} &=& (f_{\O3x}+f_{\O2x})/\sqrt{2}\;, \nn
\ft_{5x} &=& (f_{\O3x}-f_{\O2x})/\sqrt{2} \;.
\mylabel{eq:orig}
\eea
For this case we find $f_{\O2x}=f_{\O3x}$, so
$\ft_{5x}$ vanishes and $\ft_{4x}$ simplifies.  Applying this
to the $T$ tensor elements, we find
\bea
\Tt_{1,xxxx} &=& T_{\Sr,xxxx} \;,\nn
\Tt_{2,xxxx} &=& T_{\Ti,xxxx} \;,\nn
\Tt_{3,xxxx} &=& T_{\O1,xxxx} \;,\nn
\Tt_{4,xxxx} &=& \sqrt{2}\,T_{\O 2,xxxx} \;,\nn
\Tt_{5,xxxx} &=& 0 \;.
\eea
These correspond to $T_{I1111}$ elements in the notation of
the main part of the manuscript.
Similar results hold for the $Q\pone$ and $Q\pthr$ tensors:
\bea
\Qt\pone_{3,xx} &=& Q\pone_{\O1,xx} \;, \nn
\Qt\pone_{4,xx} &=& \sqrt{2}\,Q\pone_{\O 2,xx} \;, \\
\Qt\pthr_{3,xxxx} &=& Q\pthr_{\O1,xxxx} \;, \nn
\Qt\pthr_{4,xxxx} &=& \sqrt{2}\,Q\pthr_{\O 2,xxxx} \;.
\eea

If instead we consider the presence of a transverse strain gradient
$\nu_{yxx}$, we find that
$f_{\O1y}\ne f_{\O2y} \ne f_{\O3y}$ are non-zero.  Again using
the transformation rules of \equ{A} we find
\bea
\Tt_{1,yyxx} &=& T_{\Sr,yyxx} \;,\nn
\Tt_{2,yyxx} &=& T_{\Ti,yyxx} \;,\nn
\Tt_{3,yyxx} &=& T_{\O2,yyxx} \;,\nn
\Tt_{4,yyxx} &=& (T_{\O1,yyxx}+T_{\O3,yyxx})/\sqrt{2} \;,\nn
\Tt_{5,yyxx} &=& (T_{\O1,yyxx}-T_{\O3,yyxx})/\sqrt{2} \;.
\eea
Using symmetry, these correspond to the $T_{I1122}$ elements in the
notation of the main part of the manuscript.

\subsection{45$^\circ$ rotated frame}

Referring now to Fig.~\ref{fig:frame}(c-d),
we consider the 45$^\circ$ rotated geometry as in Fig.~\ref{fig:model}(b).
Then to relate the forces, we have to carry out 45$^\circ$
rotations on Cartesian indices twice, once before and once after
the transformation
to mode variables.  This is trivial for the Sr and Ti atoms,
giving $\ft_{1x'}=f_{\Sr x'}$ etc., but for the oxygens it is
more complex.  We find, for example,
\bea
\ft_{3x'} &=& (\ft_{3x}+\ft_{3y})/\sqrt{2} \;,\nn
          &=& (f_{\O1x}+f_{\O2y})/\sqrt{2} \;,\nn
          &=& (f_{\O1x'}-f_{\O1y'}+f_{\O2x'}+f_{\O2y'})/2 \;.
\eea
Using similar algebra, we find the full set of transformations
to be given by
\beq
\begin{pmatrix}
\ft_{3x'} \cr
\ft_{3y'} \cr
\ft_{4x'} \cr
\ft_{4y'} \cr
\ft_{5x'} \cr
\ft_{5y'}
\end{pmatrix}
=
\begin{pmatrix}
  h &\bar{h} & h & h & 0 & 0 \cr
 \bar{h} & h & h & h & 0 & 0 \cr
  t & t & t &\bar{t} & s & 0 \cr
  t & t &\bar{t} & t & 0 & s \cr
  t & t &\bar{t} & t & 0 &\bar{s} \cr
  t & t & t &\bar{t} & \bar{s}& 0
\end{pmatrix}
\cdot
\begin{pmatrix}
f_{\O1x'}\cr
f_{\O1y'}\cr
f_{\O2x'}\cr
f_{\O2y'}\cr
f_{\O3x'}\cr
f_{\O3y'}
\end{pmatrix}
\mylabel{eq:rot}
\eeq
where $h=1/2$, $s=\sqrt{2}$, $t=1/2\sqrt{2}$, and a
bar indicates a minus sign.

We restrict our attention
now to longitudinal strain gradients of the form $\nu_{x'x'x'}$.
For the oxygens we find
$f_{\O1x'}=f_{\O2x'}$, $f_{\O1y'}=-f_{\O2y'}$, and $f_{\O3x'}$ are
non-zero.  Then using \equ{rot} we find
\bea
\ft_{3x'} &=& f_{\O1x'}-f_{\O1y'} \nn
\ft_{4x'} &=& (f_{\O1x'}+f_{\O1y'}+f_{\O3x'})/\sqrt{2} \nn
\ft_{5y'} &=& (f_{\O1x'}+f_{\O1y'}-f_{\O3x'})/\sqrt{2}
\eea
while $\ft_{3y'}=\ft_{4y'}=\ft_{5x'}=0$.
It follows that
\bea
\Tt_{1,x'x'x'x'} &=& T_{\Sr,x'x'x'x'} \nn
\Tt_{2,x'x'x'x'} &=& T_{\Ti,x'x'x'x'} \nn
\Tt_{3,x'x'x'x'} &=& T_{\O1,x'x'x'x'}- T_{\O1,y'x'x'x'} \nn
\Tt_{4,x'x'x'x'} &=& (T_{\O1,x'x'x'x'}+T_{\O1,y'x'x'x'} \nn
        &&\hspace{0.7cm} +T_{\O3,x'x'x'x'})/\sqrt{2} \nn
\Tt_{5,y'x'x'x'} &=& (T_{\O1,x'x'x'x'}+T_{\O1,y'x'x'x'} \nn
        &&\hspace{0.7cm} -T_{\O3,x'x'x'x'})/\sqrt{2}  \;.
\eea
Similarly, for the $\Qt\pthr$ tensors in the rotated frame
we find
\bea
\Qt\pthr_{3,xxxx} &=& Q\pthr_{\O1,x'x'x'x'}-Q\pthr_{\O1,x'y'x'x'} \;,\nn
\Qt\pthr_{4,xxxx} &=& ( Q\pthr_{\O 1,x'x'x'x'} + Q\pthr_{\O 1,x'y'x'x'} \nn
   &&\hspace{0.7cm} + Q\pthr_{\O 3,x'x'x'x'})/\sqrt{2} \;,
\eea
where we used that
$Q\pthr_{\O 1,x'x'x'x'}=Q\pthr_{\O 2,x'x'x'x'}$ and
$Q\pthr_{\O 1,x'y'x'x'}=-Q\pthr_{\O 2,x'y'x'x'}$ from symmetry.

\subsection{Discussion}

Note that $\Tt_{\sigma\a\b\c\d}$ has the same symmetry for
$\sigma\in\{1,2,3,4\}$. That is, we have arranged things so that
the $\sigma$=3 and 4 cases behave just like
$\sigma$=1 (Sr) or $\sigma$=2 (Ti), so that any formulas used
for Si and Ti contributions can easily
be extended to the oxygen modes of $\GOF$ symmetry.  Note that
strain gradients also induce forces of $\GTF$ symmetry,
corresponding to $\sigma$=5, which in turn cause first-order $\GTF$
displacements.  However, because these modes are not IR-active,
they do not contribute to the flexoelectric response.


\bibliography{flexo,exper}

\begin{thebibliography}{87}
\bibitem[{exp()}]{explan-cell}
\bibinfo{note}{$V_c$ may be either the conventional or primitive cell volume,
  as long as the sum $\sum_I$ runs over the atoms contained in this volume.}
\bibitem[{not({\natexlab{a}})}]{notes-mu-g}
\bibinfo{note}{Eq.~(\ref{eq:sym-FEC}) can also be obtained more directly by
  rewriting Eq.~(\ref{eq:nu-nus}) as $\n\bbcd =
  \nb_{\b\c\d}+\nb_{\d\b\c}-\nb_{\c\d\b}$, since $\nb_{\b\c\d}$ is symmetric in
  the first two indices. Inserting this into $P_{\a} = g_{\a\d\b\c}
  \nb_{\b\c\d}$ yields Eq.~(\ref{eq:sym-FEC}) directly. However, the
  presentation in the main text gives more perspective about the possible
  choices concerning the definitions of the flexocoefficients under symmetric
  and unsymmetric strain.}
\bibitem[{not({\natexlab{b}})}]{notes-Te}
\bibinfo{note}{For \BTO, \PTO\ and \STO, using Eq.~(\ref{eq:muldT}) to
  calculate $\m\pldE$ is not practical because the force-constant matrix $K$ is
  non-physical due to the ferroelectric instability at 0\,K. We use
  Eq.~(\ref{eq:K2d-e}) to convert $T^{\E}$ to $T^D$ and calculate the full
  tensor $\mu\pldD$, which also requires $\mu\pel _{1122}$ and $\mu\pel
  _{1221}$. Then we convert it to the value at fixed $\E$.}
\bibitem[{not({\natexlab{c}})}]{notes-gpos}
\bibinfo{note}{For example, choosing $\mu\pel_{1122} = 0.088 \, \mu\pel\bltwo$,
  we obtained $g_{1111}=-36.9$\,nC/m, $g_{1221} = -1.3$\,nC/m, and $g_{1122} =
  -3.6$\,nC/m for \STO, which gives a positive $g^{\rm eff}=6.2$\,nC/m and
  agrees well with the experiment results.\cite{zubko-prl07,zubko-prl07-err}}.
\end{thebibliography}

\end{document}